\documentclass[aps,onecolumn,11pt,floatfix,altaffilletter,superscriptaddress,preprintnumbers,tightenlines,showpacs,showkeys,notitlepage,nofootinbib]{revtex4-2} 
               
\usepackage{url}

\usepackage{amsfonts,amsmath,amssymb}
\usepackage{mathtools}
\usepackage{graphicx}
\usepackage[colorlinks=true, allcolors=blue]{hyperref}
\usepackage[capitalise]{cleveref}
\usepackage{xcolor}
\usepackage{orcidlink}  
\usepackage[normalem]{ulem}

\begin{document}

\title{Maximal value for trilinear Higgs coupling in a 3-3-1 EFT}
\author{Adriano Cherchiglia \orcidlink{0000-0002-2457-1671}}
\affiliation{Instituto de Física Gleb Wataghin UNICAMP,
Rua Sérgio Buarque de Holanda, Campinas, SP, Brazil}
\author{Leonardo J. Ferreira Leite \orcidlink{0000-0002-4574-9336}}
\affiliation{Instituto de Física Gleb Wataghin UNICAMP,
Rua Sérgio Buarque de Holanda, Campinas, SP, Brazil}

\begin{abstract}
\noindent
Recent efforts, both theoretical and experimental, have increasingly focused on the scalar potential of the Standard Model, with a highlight on the trilinear Higgs coupling. This parameter has long been recognized for its potential to test Beyond-Standard-Model (BSM) theories and its significance in understanding early cosmological dynamics. In order to broadly map BSM scenarios, a powerful tool is to devise its effective field theory (EFT) version for low-energies. In this work, we obtain a consistent EFT for a class of models based on the gauge group $SU(3)_c\times SU(3)_L\times U(1)_Y$. After properly matching the UV-complete theory at one-loop, we show that the EFT is a Two-Higgs-Doublet Model (2HDM), where some of the quartic couplings are naturally small. By imposing bounds from electroweak precision observables, collider, flavor, as well as theoretical considerations, we obtain that the maximum value of the trilinear Higgs coupling is more than four times larger than the SM prediction, potentially testable at the LHC Hi-Lumi upgrade and other future colliders. Moreover, we find that such large values are only attainable if one considers an out-of-alignment scenario, even if the deviation is very small.
\end{abstract}

\maketitle

\section{Introduction}

One of the major achievements of the LHC experimental program was to find the last missing piece of the Standard Model (SM), the Higgs boson~\cite{ATLAS:2012yve, CMS:2012qbp}. After its discovery in 2012, a dedicated program started to assess whether the scalar found by the LHC collaborations was indeed the one predicted by the SM. At present, its coupling to weak gauge bosons as well as third-generation fermions has been scrutinized, with excellent agreement~\cite{ATLAS:2022vkf,CMS:2022dwd}. Soon, even the precision of its coupling to second generations fermions is expected to be increased. Even though the present scenario points to a scalar that behaves very SM-like, in particular regarding the Yukawa and kinetic Lagrangian, the scalar potential is known to a lesser extent. If the shape of the scalar potential presents deviations from the SM prediction, it could help to pave the way to Beyond Standard Model (BSM) scenarios, in particular in the case that no new mass resonances are found at the LHC. Particularly promising is the trilinear Higgs coupling $\lambda_{hhh}$, whose $\kappa$ multiplier, $\kappa_{\lambda}=\lambda_{hhh}/(\lambda_{hhh})_{SM}$, is expected to be restricted to [0.5,1.6] at 68$\%$ confidence-level in the LHC high luminosity phase (HL-LHC)~\cite{Chang:2019ncg,Cepeda:2019klc,ATL-PHYS-PUB-2022-053}.

Knowledge of the scalar potential may also help unveil baryogenesis mechanisms, which typically requires the inclusion of additional scalars in the SM~\cite{Kanemura:2004ch}. Some of the simplest extensions have been studied, where only scalar doublets and/or singlets are added; see \cite{Robens:2019kga,Abouabid:2021yvw} for a recent analysis and~\cite{Brigljevic:2024vuv} for a review. Regarding the 2HDM, even two-loop corrections to $\kappa_{\lambda}$ are known~\cite{Kanemura:2004mg,Braathen:2019pxr,Braathen:2019zoh}. Surprisingly, it was shown in~\cite{Bahl:2022jnx} that the present knowledge of the trilinear Higgs coupling already restricts a region of the 2HDM parameter space allowed by all other relevant constraints. This conclusion relies on the inclusion of two-loop corrections, which can be up to 60$\%$ larger than the one-loop corrections for some corners of the parameter space. 

The analysis for $\kappa_{\lambda}$ in models with a more complex scalar sector is less explored. Among those, particularly interesting is the 3-3-1 model~\cite{Pisano:1992bxx,Frampton:1992wt}, where the gauge group of the SM is extended to $SU(3)_C \otimes SU(3)_L \otimes U(1)_X$. Apart from providing an explanation for the number of families observed in Nature, it allows a mechanism to generate neutrino masses and mixings~~\cite{Montero:2000rh, Tully:2000kk,Montero:2001ts,Cortez:2005cp,Cogollo:2009yi,Cogollo:2010jw,Cogollo:2008zc,Dias:2012xp,Okada:2015bxa,Vien:2018otl,Nguyen:2018rlb,Pires:2018kaj,CarcamoHernandez:2019iwh,CarcamoHernandez:2019vih,CarcamoHernandez:2020pnh,Hong:2024yhk}, contain dark matter candidates \cite{Fregolente:2002nx,Hoang:2003vj,deS.Pires:2007gi,Mizukoshi:2010ky,Ruiz-Alvarez:2012nvg,Profumo:2013sca,Dong:2013ioa,Dong:2013wca,Cogollo:2014jia,Dong:2014wsa,Dong:2014esa,Kelso:2014qka,Dong:2014esa,Mambrini:2015sia,Dong:2015rka,deSPires:2016kkg,Alves:2016fqe,RodriguesdaSilva:2014gbi,Carvajal:2017gjj,Dong:2017zxo,Arcadi:2017xbo,Montero:2017yvy,Huong:2019vej,Alvarez-Salazar:2019cxw,VanLoi:2020xcq,Dutra:2021lto,Oliveira:2021gcw,Guzzo:2022beb}, and address the strong CP problem \cite{Pal:1994ba,Dias:2003zt,Dias:2003iq,Montero:2011tg,Dias:2018ddy,Dias:2020kbj}. By extending the gauge group, the 3-3-1 model contains a new set of gauge bosons, which have not been observed yet. Thus, experimental collaborations can set lower bounds on their masses. For instance, the $Z^{\prime}$ mass has at present a lower bound around 4 TeV \cite{Alves:2022hcp}. In this scenario, it is natural to consider the case where the new gauge bosons have been integrated out, rendering an effective field theory (EFT). It was first argued in~\cite{Okada:2016whh,Fan:2022dye}, and further extended in~\cite{Cherchiglia:2022zfy}, that the remaining particles, with masses at the electroweak scale, are the same as in the 2HDM~\cite{Branco:2011iw}. However, in these analyzes, only the case where the VEV of the first spontaneous symmetry breaking (from $SU(3)_C \otimes SU(3)_L \otimes U(1)_X$ to $SU(3)_C \otimes SU(2)_L \otimes U(1)_X$) was extremely large was considered, completely decoupling a set of particles whose masses are proportional to this large scale. In the present work, we provide a more consistent treatment of the underlying EFT, performing explicitly the matching between the 3-3-1 model and the 2HDM at the one-loop level. We will show that the scalar potential of 3-3-1 EFT can be entirely mapped into the most general 2HDM, with some of the quartic couplings being naturally suppressed. This will effectively render one-less free parameter for the 3-3-1 EFT, when compared not to the general 2HDM, but to the scenario most studied in which a $\mathbb{Z}_2$ symmetry is imposed at one of the doublets. Moreover, regarding the Yukawa Lagrangian, since in the 3-3-1 model the quark families are not all in the same representation, it is not possible to couple all families of each quark type (down, up) to the same Higgs doublet. Thus, necessarily, we have to treat one family distinctly, and the 3-3-1 EFT cannot be mapped to none of the usual four 2HDM types. 

After properly defining the 3-3-1 EFT, we provide a comprehensive analysis of $\kappa_{\lambda}$, with the aim of finding its maximal value after imposing a set of experimental and theoretical bounds. Given that we have one-less parameter in comparison to the 2HDM, we will find a lower maximum value than the one reported in~\cite{Bahl:2022jnx}. Nevertheless, it can still be probed at the HL-LHC~\cite{Chang:2019ncg,Cepeda:2019klc,ATL-PHYS-PUB-2022-053}. In the literature, the exact alignment condition is generally assumed for 2HDM, in order to avoid stringent constraints from colliders~\cite{ATLAS:2024lyh}. However, it is still possible to allow small deviations, which may open up a new region in parameter space that is particularly relevant for $\kappa_{\lambda}$~\cite{Arco:2020ucn,Arco:2022xum,Heinemeyer:2024vqw}. We have mapped the allowed region for 2HDM Type-I and Type-II as well as the 3-3-1 EFT, given the present bounds. Given the relevance of this region, we also extended the analytic formulas for $\kappa_{\lambda}$ at one-loop~\cite{Kanemura:2004mg} for the non-alignment case. Extending the two-loop formula~\cite{Braathen:2019pxr,Braathen:2019zoh} is beyond the scope of this work. However, given its importance for the maximum allowed value of $\kappa_{\lambda}$ in the strict alignment case~\cite{Bahl:2022jnx}, this enterprise may be particularly promising, which we leave for future investigation. 

This work is organized as follows: in~\cref{sec:model} we introduce the effective field description of the 3-3-1 model. In~\cref{sec:h3} we discuss the trilinear Higgs coupling ($\lambda_{hhh}$) in detail, comparing the differences expected from the 3-3-1 EFT and the 2HDM.~\cref{sec:pheno} is devoted to the description of the phenomenological constraints our model is subjected to. We also perform a numerical evaluation of the available parameter space of our model, with emphasis on the maximum value allowed for $\lambda_{hhh}$. We conclude in~\cref{sec:conclusions}. We provide three appendices. \cref{ap:2HDM} contains an detailed analysis for $\kappa_{\lambda}$ on Types I and II of 2HDM, while \cref{ap:two-loop} is related to the impact of including two-loop corrections for $\lambda_{hhh}$. Finally, in \cref{ap:one-loop} we collect the one-loop matching results for dimension-four coefficients as well as dimension-six terms induced by the heavy spectrum.

\section{The 3-3-1 EFT model}
\label{sec:model}

The SM can be extended in a number of ways. The simplest proposal consists of the addition of new particles grouped in irreducible representations of the SM gauge group. This approach is mainly guided by phenomenology, where some new phenomena (dark matter, neutrino masses, and mixing, for instance) are explained due to the interaction of these new particles with themselves and at least some of their SM counterparts.

Another proposal consists in the modification of the SM gauge group. Models in this category usually replace the SM gauge group by one single larger group, in the hope of providing a unification of the SM gauge couplings. In the same category, less ambitious models replace just one of the SM gauge sub-groups (or add new sub-groups). The 3-3-1 model fits this scenario, where the SM gauge group is replaced by $SU(3)_{c}\times SU(3)_{L} \times U(1)_{X}$. In order to reproduce the SM particles and interactions, a set of electroweak symmetry breakings (EWSB) are proposed. We will consider a 3-3-1 version where only three scalar triplets $\eta$, $\rho$, and $\chi$ are added, whose VEVs are responsible for the masses of gauge bosons as well as charged fermions. In accordance with standard notation, the first gauge breaking will be due to the VEV of $\chi$, while the second breaking comes from the VEV of both $\eta$, $\rho$. Schematically,
\begin{equation*}
SU(3)_C \otimes SU(3)_L \otimes U(1)_X \xRightarrow{v_{\chi}} SU(3)_C \otimes SU(2)_L \otimes U(1)_X \xRightarrow{v_{\eta},v_{\rho}} SU(3)_C \otimes U(1)_{\text{em}}
\end{equation*}

In other to set our notation, we will choose the scalar triplets to have the following irreducible representation under the $SU(3)_{c}\times SU(3)_{L} \times U(1)_{X}$ gauge group,
\begin{small}
    \begin{equation}
    \eta=\begin{pmatrix}
    \eta^0 \\ \eta^- \\ \eta^{-A}
    \end{pmatrix}\sim (1, \textbf{3}, X_{\eta}),
    \quad
    \rho = \begin{pmatrix}
    \rho^+ \\ \rho^0 \\ \rho^{-B}  
    \end{pmatrix}\sim (1, \textbf{3}, X_{\rho}), 
    \quad
    \chi = \begin{pmatrix}
    \chi^A \\ \chi^{B} \\ \chi^0
    \end{pmatrix}\sim (1, \textbf{3}, X_{\chi}),
\end{equation}
\end{small}
where the charges under $U(1)_X$ are given by
\begin{equation}
    X_{\eta}=-\frac{1}{2} -\frac{\beta_{Q}}{2\sqrt{3}},\quad X_{\rho}=\frac{1}{2}-\frac{\beta_{Q}}{2\sqrt{3}}, \quad X_{\chi}=\frac{\beta_{Q}}{\sqrt{3}}
\end{equation}
Notice that we have some freedom on the choice of the parameter $\beta_{Q}$. In terms of it, the electric charges of particles of type $A$ or $B$ are
\begin{equation}
Q^{A} = \frac{1}{2}+\frac{\sqrt{3}}{2}\beta_{Q},\quad Q^{B} = -\frac{1}{2}+\frac{\sqrt{3}}{2}\beta_{Q}
\end{equation}

In terms of the scalar triplets, the scalar potential takes the form
\begin{align}
V\left(\eta,\rho,\chi\right) = &  \mu_{1}^{2}\rho^{\dagger}\rho+\mu_{2}^{2}\eta^{\dagger}\eta+\mu_{3}^{2}\chi^{\dagger}\chi+\lambda_{1}\left(\rho^{\dagger}\rho\right)^{2}+\lambda_{2}\left(\eta^{\dagger}\eta\right)^{2}+\lambda_{3}\left(\chi^{\dagger}\chi\right)^{2} \nonumber\\
  & +\lambda_{12}\left(\rho^{\dagger}\rho\right)\left(\eta^{\dagger}\eta\right)+\lambda_{13}\left(\chi^{\dagger}\chi\right)\left(\rho^{\dagger}\rho\right)+\lambda_{23}\left(\eta^{\dagger}\eta\right)\left(\chi^{\dagger}\chi\right)\nonumber \\
  & +\zeta_{12}\left(\rho^{\dagger}\eta\right)\left(\eta^{\dagger}\rho\right)+\zeta_{13}\left(\rho^{\dagger}\chi\right)\left(\chi^{\dagger}\rho\right)+\zeta_{23}\left(\eta^{\dagger}\chi\right)\left(\chi^{\dagger}\eta\right)\nonumber\\
  & -\sqrt{2}f\epsilon_{ijk}\eta_{i}\rho_{j}\chi_{k}+\textrm{h.c.},
 \label{eq:Vscalar}
\end{align}

In the equation above, we implicitly assume that $\beta_{Q}\neq \pm 1/\sqrt{3}$, otherwise terms with an odd number of fields would be present. Notice, however, that the case $\beta_{Q}= \pm 1/\sqrt{3}$ can still be considered if a $\mathbb{Z}_2$ symmetry (under which only $\chi$ is odd) is implicitly assumed\footnote{The $\mathbb{Z}_2$ symmetry will be softly broken by the term $\sqrt{2}f\epsilon_{ijk}\eta_{i}\rho_{j}\chi_{k}$. If this term is discarded, one generates a massless CP-odd scalar, which is disfavored by experiment}. Once $\chi^{0}$ acquires a VEV, the remaining symmetry is $SU(3)_{c}\times SU(2)_{L} \times U(1)_{Y}$. At this stage, we can group the scalar fields in the following representations under the SM gauge group,
\begin{align}
\Phi_\eta=\begin{pmatrix}
\eta^0\\ -\eta^-
\end{pmatrix}
\sim (1,2,-1/2),
\quad
\Phi_\rho=\begin{pmatrix}
\rho^+\\ \rho^0
\end{pmatrix}\sim (1,2,1/2),
\quad
\Phi_\chi=\begin{pmatrix}
\chi^{A}\\ \chi^{B}
\end{pmatrix}\sim (1,2,\sqrt{3}\beta_{Q}/2)\nonumber\\
\eta^{-A}\sim(1,1,-Q^{A}),
\quad
\rho^{-B}\sim(1,1,-Q^{B}),
\quad
\chi^{0}\sim(1,1,0)    ,
\end{align}
while the scalar potential can be rewritten as

\begin{align}
V\left(\eta,\rho,\chi\right)  &=   \mu_{1}^{2}(\Phi_{\rho}^{\dagger}\Phi_{\rho}+\rho^{B}\rho^{-B})+\mu_{2}^{2}(\Phi_{\eta}^{\dagger}\Phi_{\eta}+\eta^{A}\eta^{-A})+\mu_{3}^{2}(\chi^{\dagger}\chi)\nonumber\\
&
+\lambda_{1}(\Phi_{\rho}^{\dagger}\Phi_{\rho}+\rho^{B}\rho^{-B})^{2}+\lambda_{2}(\Phi_{\eta}^{\dagger}\Phi_{\eta}+\eta^{A}\eta^{-A})^{2}+\lambda_{3}\left(\chi^{\dagger}\chi\right)^{2} \nonumber\\
   & +\lambda_{12}(\Phi_{\rho}^{\dagger}\Phi_{\rho}+\rho^{B}\rho^{-B})(\Phi_{\eta}^{\dagger}\Phi_{\eta}+\eta^{A}\eta^{-A})\nonumber\\
  &
+\lambda_{13}\left(\chi^{\dagger}\chi\right)(\Phi_{\rho}^{\dagger}\Phi_{\rho}+\rho^{B}\rho^{-B})+\lambda_{23}(\Phi_{\eta}^{\dagger}\Phi_{\eta}+\eta^{A}\eta^{-A})\left(\chi^{\dagger}\chi\right)\nonumber \\
   & +\zeta_{12}\left(\Phi_{\rho}^{\dagger}\Phi_{\eta}+\rho^{B}\eta^{-A} \right)\left(\Phi_{\eta}^{\dagger}\Phi_{\rho}+\eta^{A}\rho^{-B} \right)\nonumber\\
  &
+\zeta_{13}\left(\Phi_{\rho}^{\dagger}\Phi_{\chi}+\rho^{B}\chi^{0}\right)\left(\Phi_{\chi}^{\dagger}\Phi_{\rho}+\chi^{0*}\rho^{-B}\right)\nonumber\\
  &
+\zeta_{23}\left(\Phi_{\eta}^{\dagger}\Phi_{\chi}+\eta^{A}\chi^{0}\right)\left(\Phi_{\chi}^{\dagger}\Phi_{\eta}+\chi^{0*}\eta^{-A}\right)\nonumber\\
   & -\sqrt{2}f(\Phi_{\eta}^{T}\epsilon\Phi_{\rho}\chi^{0} + \Phi_{\rho}^{\dagger}\Phi_{\chi}\rho^{-B} + \Phi^{\dagger}_{\eta}\Phi_{\chi}\eta^{-A}
 +\textrm{h.c.})\,.\label{eq:Vscalar2}
\end{align}

It should be noticed that, in general, the scalar fields are not mass eigenstates yet. We recall that the present limit on the extra gauge bosons introduced by the 3-3-1 model is around 4 TeV, which sets a lower bound on $v_{\chi}\sim10$ TeV. Thus, it is reasonable to assume $v_{\chi} \gg v_{\rho},v_{\eta}$. In this scenario, the mixing between $\chi_{0}$ and the neutral components of $\Phi_{\rho,\eta}$ is suppressed~\cite{Okada:2016whh,Fan:2022dye}, rendering $\chi_{0}$ a mass eingenstate. The fields $\eta^{A},\rho^{B}$, on the other hand, are already mass eingenstates. As we are going to show, for all three fields ($\chi_{0}$, $\eta^{A}$, $\rho^{B}$) their masses are proportional to $v_{\chi}$ in the limit $v_{\chi}\gg v_{\rho},v_{\eta}$. Thus, these scalars can be considered heavy, in comparison to $\Phi_{\rho},\Phi_{\eta}$, which will contain light fields. Defining $\left< \chi_0 \right>=v_{\chi}$, we obtain the tadpole equation,
\begin{align}\label{eq:tadpole}
   \frac{1}{v_{\chi}}\frac{\partial V}{\partial \chi_{0}}\Big|_{\chi_{0}=v_{\chi}} =   2\mu_{3}^{2} + 4 \lambda_{3}v_{\chi}^{2} \rightarrow \mu_{3}^{2} = - 2 \lambda_{3}v_{\chi}^{2} ,
\end{align}
as well as the masses for the heavy fields ($\Phi_{\chi}$ contains the Goldstones of the new gauge bosons),
\begin{align}
   m_{\chi_{0}}^{2} = \frac{\partial^{2} V}{\partial \chi_{0}^{2}}\Big|_{\chi_{0}=v_{\chi}} =&\;  2\mu_{3}^{2} + 12 \lambda_{3}v_{\chi}^{2} = 8 \lambda_{3}v_{\chi}^{2}\,,
\end{align}
\begin{align}
   m_{\rho_{B}}^{2}=\frac{\partial^{2} V}{\partial \rho^{B}\rho^{-B}}\Big|_{\chi_{0}=v_{\chi}} =\;  \mu_{1}^{2} + \zeta_{13}v_{\chi}^{2}\,, \quad\quad 
   m_{\eta_{A}}^{2}=\frac{\partial^{2} V}{\partial \eta^{A}\eta^{-A}}\Big|_{\chi_{0}=v_{\chi}} =\;  \mu_{2}^{2} + \zeta_{23}v_{\chi}^{2} \,.
\end{align}

The parameters $\mu_{1,2}^{2}$ can be removed when considering the second gauge breaking. However, since the scalar potential has only four dimensionful parameters ($\mu_{1,2,3}^{2}$, and $f$), and after both breakings we have three VEVs, we can trade $\mu_{1,2,3}^{2}$ for the VEVs, in similarity to~\cref{eq:tadpole}. It is easy to show that $\mu^{2}_{1,2}\sim v_{\eta,\rho}^{2}$, which implies that in the limit $v_{\chi} \gg v_{\rho},v_{\eta}$, all heavy masses are proportional to $v_{\chi}$, as previously stated. 

In order to define a sensible effective field theory, we will integrate out the heavy fields $\chi,\;\rho_{B},\;\eta_{A}$. At tree-level matching, this can be conveniently done by solving the equation of motion (E.O.M.) for the heavy fields, defining our EFT Lagrangian. Notice that the fields $\Phi_{\chi}$, $\rho_{B},\;\eta_{A}$ appear only in pairs in~\cref{eq:Vscalar}. Thus, considering tree-level matching, they can be integrated out without any consequence at low-energy. The only exception is for terms containing a single $\chi^{0}$, for instance ($\Phi_{\eta}^{T}\epsilon\Phi_{\rho}\chi^{0}$), where there is just one heavy field coupled to light ones. Thus, as long as we consider only tree-level matching, terms of this kind are the only relevant. 
The EOM for $\chi_{0}$ can be formally solved by
\begin{align}
   \chi_{0} \approx & \frac{\sqrt{2}f\left(\Phi_{\eta}^{T}\epsilon\Phi_{\rho} + \textrm{h.c.}\right)}{m^{2}_{\chi_{0}}} - \frac{2\lambda_{13}v_{\chi}\Phi_{\rho}^{\dagger}\Phi_{\rho}}{m^{2}_{\chi_{0}}} - \frac{2\lambda_{23}v_{\chi}\Phi_{\eta}^{\dagger}\Phi_{\eta}}{m^{2}_{\chi_{0}}}\,,
\end{align}
where we implicitly assume that $f \ll v_{\chi}$. Finally, the scalar potential in the EFT, containing terms up to dimension 4, is given by
\begin{eqnarray}
V\left(\eta,\rho,\chi\right) & = &  (\mu_{1}^{2} + \lambda_{13}v_{\chi}^{2})(\Phi_{\rho}^{\dagger}\Phi_{\rho})+(\mu_{2}^{2}+\lambda_{23}v_{\chi}^{2})(\Phi_{\eta}^{\dagger}\Phi_{\eta})\nonumber\\
& & + \left[\lambda_{1}-\frac{2v_{\chi}^{2}\lambda_{13}^{2}}{m_{\chi_{0}}^{2}}\right](\Phi_{\rho}^{\dagger}\Phi_{\rho})^{2}+\left[\lambda_{2}-\frac{2v_{\chi}^{2}\lambda_{23}^{2}}{m_{\chi_{0}}^{2}}\right](\Phi_{\eta}^{\dagger}\Phi_{\eta})^{2} 
 \nonumber\\
 &  & +\left[\lambda_{12}-\frac{4v_{\chi}^{2}\lambda_{13}\lambda_{23}}{m_{\chi_{0}}^{2}}\right](\Phi_{\rho}^{\dagger}\Phi_{\rho})(\Phi_{\eta}^{\dagger}\Phi_{\eta})+\left[\zeta_{12}-\left(\frac{2f^{2}}{m^{2}_{\chi_{0}}}\right)\right](\Phi^T_\eta\varepsilon\Phi_\rho)^\dagger(\Phi_\eta^T\varepsilon\Phi_\rho)\nonumber\\
&& + \Bigg[-\sqrt{2}fv_{\chi}(\Phi_{\eta}^{T}\epsilon\Phi_{\rho})   
 -\left(\frac{f^{2}}{m^{2}_{\chi_{0}}}\right)(\Phi_{\eta}^{T}\epsilon\Phi_{\rho})^{2}
 +\left(\frac{2\sqrt{2}fv_{\chi}\lambda_{13}}{m^{2}_{\chi_{0}}}\right)\left(\Phi_{\eta}^{T}\epsilon\Phi_{\rho}\right)(\Phi_{\rho}^{\dagger}\Phi_{\rho}) \nonumber\\
&&
 +\left(\frac{2\sqrt{2}fv_{\chi}\lambda_{23}}{m^{2}_{\chi_{0}}}\right)(\Phi_{\eta}^{\dagger}\Phi_{\eta})\left(\Phi_{\eta}^{T}\epsilon\Phi_{\rho}\right) + \textrm{h.c.} \Bigg].
\end{eqnarray}

It is immediate to notice that the above equation can be entirely mapped into the scalar potential of the 2HDM 
\begin{align}
\label{eq:Vscalar2HDM}
V\left(\Phi_{1},\Phi_{2}\right)  = &   m_{11}^{2}\Phi_{1}^{\dagger}\Phi_{1}+m_{22}^{2}\Phi_{2}^{\dagger}\Phi_{2}+\frac{\Lambda_{1}}{2}\left(\Phi_{1}^{\dagger}\Phi_{1}\right)^{2}+\frac{\Lambda_{2}}{2}\left(\Phi_{2}^{\dagger}\Phi_{2}\right)^{2} \nonumber\\
 & +\Lambda_{3}\left(\Phi_{1}^{\dagger}\Phi_{1}\right)\left(\Phi_{2}^{\dagger}\Phi_{2}\right)+\Lambda_{4}\left(\Phi_{1}^{\dagger}\Phi_{2}\right)\left(\Phi_{2}^{\dagger}\Phi_{1}\right) + \big[-m_{12}^{2}\Phi_{1}^{\dagger}\Phi_{2} \nonumber\\ &+ \frac{\Lambda_{5}}{2}\left(\Phi_{1}^{\dagger}\Phi_{2}\right)^{2} + \Lambda_{6}\left(\Phi_{1}^{\dagger}\Phi_{1}\right)\left(\Phi_{1}^{\dagger}\Phi_{2}\right)+\Lambda_{7}\left(\Phi_{2}^{\dagger}\Phi_{2}\right)\left(\Phi_{1}^{\dagger}\Phi_{2}\right)+\textrm{h.c.}\big],
\end{align}
by performing the mapping $\Phi_{\rho} \rightarrow \Phi_{1}$ and $\Phi_{\eta}^{\tiny{C}} \rightarrow \Phi_{2}$. Explicitly, the tree-level matching between the scalar sector of the 3-3-1 model and the 2HDM is given by
\begin{align}
\Lambda_{1}&= 2\left[\lambda_{1}-\frac{2v_{\chi}^{2}\lambda_{13}^{2}}{m_{\chi_{0}}^{2}}\right],\;\quad\Lambda_{2}=2\left[\lambda_{2}-\frac{2v_{\chi}^{2}\lambda_{23}^{2}}{m_{\chi_{0}}^{2}}\right],\;\quad\Lambda_{3}=\left[\lambda_{12}-\frac{4v_{\chi}^{2}\lambda_{13}\lambda_{23}}{m_{\chi_{0}}^{2}}\right],\;\nonumber\\
\Lambda_{4}&=\left[\zeta_{12}-\left(\frac{2f^{2}}{m^{2}_{\chi_{0}}}\right)\right],\;\quad\Lambda_{5}=-\frac{2f^{2}}{m^{2}_{\chi_{0}}},\;\quad\Lambda_{6}=\frac{2\sqrt{2}fv_{\chi}\lambda_{13}}{m^{2}_{\chi_{0}}},\;\quad\Lambda_{7}=\frac{2\sqrt{2}fv_{\chi}\lambda_{23}}{m^{2}_{\chi_{0}}}
\end{align}

Some comments are in order: when obtaining the masses of the scalars (in the limit $v_{\chi} \gg v_{\rho},v_{\eta}$), one finds the relation $f v_{\chi}\sim m_{A}^{2}s_{\beta}c_{\beta}$, where $A$ is the pseudo-scalar, and $\tan\beta = v_{\eta}/v_{\rho}$. Thus, the scale $f$ is approximately given by $f\sim m_{A}^{2}/v_{\chi}$. By assuming that $m_{A} \ll v_{\chi}$, it is immediate to see that the 3-3-1 EFT scalar potential has $\Lambda_{5}$ suppressed. Another relation valid in the limit $v_{\chi} \gg v_{\rho},v_{\eta}$ is $\lambda_{i3}\sim m_{A}^{2}/v_{\chi}^2$. Thus, we find that in the 3-3-1 EFT $\Lambda_{6,7}$ are also suppressed. The suppression of $\Lambda_{5,6,7}$ can also be explained in more general terms. Consider the scalar potential after the first gauge breaking, \cref{eq:Vscalar2}. By removing the Goldstones related to the new charged gauge bosons, field $\Phi_{\chi}$, the potential possesses a $\mathbb{Z}_{n}$ symmetry in either $\Phi_{\eta}$ or $\Phi_{\rho}$, apart from soft-broken terms proportional to $f$. Thus, by suppressing the $f$ scale, and retaining only the fields $\Phi_{\eta,\rho}$ at low-scale (compared to $v_{\chi}$), one must obtain a scalar potential in the EFT that also possesses a $\mathbb{Z}_{n}$ symmetry on $\Phi_{\eta}$ or $\Phi_{\rho}$. However, notice that this is not the case for general 2HDM, \cref{eq:Vscalar2HDM}. In the latter, apart from the soft-breaking term (proportional to $m_{12}^{2}$), there are quartic terms that break the symmetry $\mathbb{Z}_{2}$ ($\Lambda_{6,7}$) or $\mathbb{Z}_{n>2}$ ($\Lambda_5$). Therefore, not only $\Lambda_{5,6,7}$ are suppressed when matching the 3-3-1 model to the 2HDM, but they are RGE-protected within the scalar sector (as long as one stays in the validity regime of the EFT). 

Up to this point, we only considered the scalar sector of the 3-3-1 when defining the 3-3-1 EFT. The influence of the heavy gauge sector can also be straightforwardly taken into account. It is achieved by resorting to the field $\Phi_{\chi}$, which contains the Goldstones of the extra gauge fields. As noted previously, these fields appear only in pairs, together with $\eta_{A}$ and $\rho_{B}$. Thus, at tree-level matching, the heavy gauge fields leave no imprint in the low-energy scalars phenomenology. We now move on to the Yukawa sector. 

When defining the fermionic fields, we have some freedom regarding which of them will transform as (anti)triplets under the 3-3-1 gauge group. In order to easily map the third quark family to a type II 2HDM, we will adopt an opposite convention to the one employed in~\cite{Cherchiglia:2022zfy}, as below
\begin{small}
    \begin{equation}
    L_{i}=\begin{pmatrix}
     \nu_{i} \\ e_{i} \\ E_{i} ,
    \end{pmatrix}_{L}\sim \left(1,3,-\frac{1}{2}-\frac{\beta_{Q}}{2\sqrt{3}}\right),
    \quad\quad
    q_{1}=\begin{pmatrix}
     d \\ -u \\ D 
    \end{pmatrix}_{L}\sim \left(3,\bar{3},\frac{1}{6}+\frac{\beta_{Q}}{2\sqrt{3}}\right),
    \end{equation}
    \begin{equation}
    q_{2}=\begin{pmatrix}
     s \\ -c \\ S 
    \end{pmatrix}_{L}\sim \left(3,\bar{3},\frac{1}{6}+\frac{\beta_{Q}}{2\sqrt{3}}\right),
     \quad\quad
    q_{3}=\begin{pmatrix}
     t \\ b \\ T 
    \end{pmatrix}_{L}\sim \left(3,3,\frac{1}{6}-\frac{\beta_{Q}}{2\sqrt{3}}\right),
\end{equation}
\begin{equation}
    u_{R},\, c_{R}\,, t_{R} \sim (3,1,2/3)\quad d_{R}\,, s_{R}\,, b_{R} \sim (3,1,-1/3), \quad\quad e_{R} \sim (1,1,-1),
\end{equation}
\begin{equation}
    D_{R}, S_{R} \sim \left(3,1,\frac{1}{6}+\frac{\sqrt{3}\beta_{Q}}{2}\right),\quad T_{R} \sim \left(3,1,\frac{1}{6}-\frac{\sqrt{3}\beta_{Q}}{2}\right), \quad E_{R} \left(1,1,-\frac{1}{2}-\frac{\sqrt{3}\beta_{Q}}{2}\right)
\end{equation}
\end{small}
 The electric charges of the $D,S,T$ fields (both left- and right-handed) are given by
\begin{equation} 
    Q_{E_{i}} = -\frac{1}{2}+\frac{\sqrt{3}\beta_{Q}}{2}\quad\quad
    Q_{D,S} = \frac{1}{6}-\frac{\sqrt{3}\beta_{Q}}{2}\quad\quad
    Q_{T} = \frac{1}{6}+\frac{\sqrt{3}\beta_{Q}}{2}
\end{equation}
Under the 3-3-1 gauge group, we can write the Yukawa lagrangian as below\footnote{For the choice $\beta_{Q}= \pm 1/\sqrt{3}$, we will consider the fields $D_{R}$, $S_{R}$, $T_{R}$, and $E_{R}$ to be odd under the $\mathbb{Z}_2$ introduced for this case.}
\begin{align}
-\mathcal{L}_{\rm{Yuk}}^{q} &= y_{ij}^{u}\bar{q}_{iL}\rho^{*} u_{jR} + y_{3j}^{u}\bar{q}_{3L}\eta u_{jR} + y_{ij}^{d}\bar{q}_{iL}\eta^{*} d_{jR} + y_{3j}^{d}\bar{q}_{3L}\rho d_{jR} \nonumber\\&+ y_{i}^{D}\bar{q}_{iL}\chi^{*} D_{R} + y_{i}^{S}\bar{q}_{iL}\chi^{*} S_{R} +y^{T}\bar{q}_{3L}\chi T_{R} + \textrm{h.c.}\\
-\mathcal{L}_{\rm{Yuk}}^{l} &= y_{mn}^{e}\bar{L}_{mL}\rho e_{nR} + y_{mn}^{E}\bar{L}_{mL}\chi E_{nR}+ \textrm{h.c.}
\end{align}
where $i=\{1,2\}$, while the other indexes go from 1 to 3. Notice that we are singling out the third-family quarks, choosing them to be components of a triplet (see, for instance~\cite{Doff:2024hid} for other possibilities). After the first gauge breaking, we obtain
\begin{align}
-\mathcal{L}_{\rm{Yuk}}^{q} &= y_{ij}^{u}\bar{q}_{iL}\Phi_{\rho}^{*}u_{jR} +y_{1j}^{u}\bar{D}_{L}\rho^{B}u_{jR} +y_{2j}^{u}\bar{S}_{L}\rho^{B}u_{jR}
+y_{3j}^{u}\bar{q}_{3L}\Phi_{\eta}u_{jR}
+y_{3j}^{u}\bar{T}_{L}\eta^{-A}u_{jR}
\nonumber\\&
+ y_{ij}^{d}\bar{q}_{iL}\Phi_{\eta}^{*}d_{jR} 
+ y_{1j}^{d}\bar{D}_{L}\eta^{A}d_{jR}
+ y_{2j}^{d}\bar{S}_{L}\eta^{A}d_{jR}
 +y_{3j}^{d}\bar{q}_{3L}\Phi_{\rho} d_{jR}
 +y_{3j}^{d}\bar{T}_{L}\rho^{-B}d_{jR}
 \nonumber\\&
 + y_{1}^{D}\bar{D}_{L}\chi_{0}^{*} D_{R}
 + y_{2}^{D}\bar{S}_{L}\chi_{0}^{*} D_{R}
 + y_{1}^{S}\bar{D}_{L}\chi_{0}^{*} S_{R}
 + y_{2}^{S}\bar{S}_{L}\chi_{0}^{*} S_{R}
 + y^{T}\bar{T}_{L}\chi_{0} T_{R}
  \nonumber\\&
 + v_{\chi}y_{1}^{D}\bar{D}_{L} D_{R}
 + v_{\chi}y_{2}^{D}\bar{S}_{L} D_{R}
 + v_{\chi}y_{1}^{S}\bar{D}_{L} S_{R}
 + v_{\chi}y_{2}^{S}\bar{S}_{L}S_{R}
 + v_{\chi}y^{T}\bar{T}_{L} T_{R} + \textrm{h.c.}\label{eq:fermionQ}
 \\
 -\mathcal{L}_{\rm{Yuk}}^{l} &= y_{mn}^{e}\bar{l}_{mL}\Phi_{\rho} e_{nR} + y_{mn}^{e}\bar{E}_{mL}\rho^{-B} e_{nR} + y_{mn}^{E}\bar{E}_{mL}\chi_{0}E_{nR}+ v_{\chi} y_{mn}^{E}\bar{E}_{mL}E_{nR}+ \textrm{h.c.}\label{eq:fermionE}
\end{align}

Considering that the yukawas ($y^{D},y^{S},y^{T},y^{E})$ are of order 1, the masses of the extra fermions will be around the scale $v_{\chi}$, which is heavy. Thus, we can also integrate them from our theory. Notice, however, that most terms contain an even number of heavy fields, the only exception being the ones containing $\chi_0$, which is odd but contains three heavy fields. Therefore, we can only have corrections at one-loop matching, and the 3-3-1 EFT Yukawa sector is simply given by

\begin{align}\label{eq:lag}
-\mathcal{L}_{\rm{Yuk}}^{q} &= y_{1j}^{u}\left(\bar{d}_{L}\rho^{-}-\bar{u}_{L}\rho^{0}\right) u_{jR} + y_{2j}^{u}\left(\bar{s}_{L}\rho^{-}-\bar{c}_{L}\rho^{0}\right) u_{jR} + y_{3j}^{u}\left(\bar{t}_{L}\eta^{0}-\bar{b}_{L}\eta^{-}\right) u_{jR}\nonumber\\ 
&+ y_{1j}^{d}\left(\bar{u}_{L}\eta^{+}+\bar{d}_{L}\eta^{0}\right) d_{jR} + + y_{2j}^{d}\left(\bar{c}_{L}\eta^{+}+\bar{s}_{L}\eta^{0}\right) d_{jR} + y_{3j}^{d}\left(\bar{b}_{L}\rho^{0}+\bar{t}_{L}\rho^{+}\right) d_{jR}  + \textrm{h.c.}\\
-\mathcal{L}_{\rm{Yuk}}^{l} &= y_{mn}^{e}\left(\bar{e}_{mL}\rho^{0}+\bar{\nu}_{mL}\rho^{+}\right)e_{nR}+ \textrm{h.c.}
\end{align} 

Resorting, for simplicity, to the third family only case, we obtain
\begin{align}\label{eq:lag_third}
-\mathcal{L}_{\rm{Yuk}}^{q} &\supset  y_{33}^{u}\left(\bar{t}_{L}\eta^{0}-\bar{b}_{L}\eta^{-}\right) t_{R} + y_{33}^{d}\left(\bar{b}_{L}\rho^{0}+\bar{t}_{L}\rho^{+}\right) b_{R}  + \textrm{h.c.}\\
-\mathcal{L}_{\rm{Yuk}}^{l} &\supset y_{33}^{e}\left(\bar{\tau}_{L}\rho^{0}+\bar{\nu}_{\tau L}\rho^{+}\right)\tau_{R}+ \textrm{h.c.}
\end{align}
Notice that the tau as well as the bottom will receive their masses through $v_{\rho}$, while the top mass comes from $v_{\eta}$. Recalling that $\tan \beta = v_{\eta}/v_{\rho}$, it is clear that $y_{33}^{u}=(m_{t}/v)(\sqrt{2}/s_{\beta})$, while $y_{33}^{d}=(m_{b}/v)(\sqrt{2}/c_{\beta})$, $y_{33}^{e}=(m_{\tau}/v)(\sqrt{2}/c_{\beta})$ where $v=\sqrt{v_{\eta}^2+v_{\rho}^2}=246$ GeV. Requiring perturbativity of the Yukawa couplings, it follows that $\tan\beta$ must be larger than unity. Moreover, the pattern for the couplings between the fermions and the scalars mimics that found in the Type II 2HDM. Thus, as long as we focus only on the third family fermions, the 3-3-1 EFT Yukawa sector can be mapped to the Type II 2HDM.

Before proceeding, we would like to summarize our approach so far. We have considered the 3-3-1 model as our UV theory, whose first gauge breaking occurred at a heavy scale compared to the EW scale. A set of new gauge bosons are induced by the model, whose masses are proportional to the heavy scale. Thus, by integrating them out, we retain only the gauge bosons already present in the SM. Regarding scalars fields, we obtain a set of doublet and singlets under the SM gauge group. The scalar potential of the 3-3-1 conveys a dimensional parameter ($f$), which, in principle, can define another scale. 

In \cite{Pinheiro:2022bcs} it was shown how different values of $f$ generate different spectrum for the scalar particles. We are interested in the case where a light scalar spectrum (of order of the EW scale) can be generated, so we will choose $f v_{\chi} \sim v_{EW}^{2}$. In this scenario, by integrating out all scalar fields whose masses are proportional to $v_{\chi}$, only two scalar doublets remain at the EW scale, whose quantum numbers are exactly the ones of the scalars doublets of the 2HDM. Finally, regarding the fermionic content, on top of the SM fields, the 3-3-1 predicts additional vector-like fermions. Those have masses proportional to $v_{\chi}$. By assuming order one Yukawas, these extra fermionic particles can be integrated out as well. Thus, the theory at the EW scale contains the same gauge bosons and fermions as in the SM, with three neutral scalars and two singly charged scalars. This is the same particle content of the 2HDM, so it is natural to us to consider the matching between the 3-3-1 model and the 2HDM EFT~\cite{Crivellin:2016ihg,Karmakar:2017yek,Anisha:2019nzx,Dermisek:2024ohe}. 

Regarding the 2HDM EFT, there are disagreements related to the number of independent dimension-six operators, as extensively discussed in~\cite{Dermisek:2024ohe}. In the latter, the operators are defined in the Higgs basis as well, not only in the symmetry basis. For our purposes, we will consider only dimension-six operators in the symmetry basis, in accordance with the potential defined in \cref{eq:Vscalar2HDM}. The basis of dimension-six operators relevant for our purposes is given in \cref{ap:one-loop}. 

In this work, we are mainly interested in the corrections to the trilinear Higgs coupling, in the context of the 3-3-1 model matched to the 2HDM EFT. In this case, as we estimate in \cref{sec:pheno}, the inclusion of dimension-six operators is a percent correction. On the other hand, from a purely 2HDM EFT perspective, the computation of the trilinear Higgs coupling is a promising research avenue, which may involve not only operators containing six scalar fields, but also others containing derivatives due to field redefinitions, in similarity to SMEFT~\cite{Alasfar:2023xpc}. However, such analysis is beyond the scope of our present work.

Up to this point, we have only considered tree-level matching. Since we aim to compute the trilinear Higgs coupling, whose main contributions are loop-induced, we need to extend the matching to one-loop for consistency. In this case, the complexity of the calculation is substantial, which requires the usage of dedicated codes such as \textit{Matchete}~\cite{FuentesMartin:2022jrf}. We have implemented the model and performed the one-loop matching for $\Lambda_{i}$ which we discuss in \cref{ap:one-loop}. As we briefly discussed, another interesting aspect of an EFT is the appearance of dimension-six (and higher) operators. We have also computed them at one-loop matching, which we discuss in \cref{ap:one-loop}.

\section{The trilinear Higgs coupling}
\label{sec:h3}

One of the main results of our contribution is to unveil the maximum allowed value for the trilinear Higgs coupling ($\lambda_{hhh}$) in the context of the 3-3-1 EFT. Before delving into this task, we will present the analytical formula for such quantity in the 2HDM up to one-loop order. At tree-level, we obtain~\cite{Kanemura:2002vm}
\begin{align}
    \lambda_{hhh}^{(0)} = &- \frac{3 m_h^2}{2vs_{2\beta}}\left[\cos(3\alpha-\beta)+3\cos(\alpha+\beta)-4\cos^{2}(\alpha-\beta)\cos(\alpha+\beta)\frac{M^{2}}{m_{h}^{2}}\right]\,,
\end{align}
where $M^{2}=m_{12}^{2}/s_{\beta}c_{\beta}$, and we have the tree-level relation
\begin{equation} 
M^{2}=m_{A}^{2}+\Lambda_{5}v^{2}+\frac{1}{2}\Lambda_{6}t_{\beta}^{-1}+\frac{1}{2}\Lambda_{7}t_{\beta}\,.
\end{equation}

The result in the SM is obtained in the alignment limit where $\sin(\beta-\alpha)=1$,
\begin{align}
    (\lambda_{hhh})_{SM} = &- \frac{3 m_h^2}{v}\,.
\end{align}
It allows us to define the $\kappa$ multiplier
\begin{align}
    \kappa_{\lambda}=\frac{\lambda_{hhh}}{(\lambda_{hhh})_{SM}}\,,
\end{align}
which we shall use hereafter. Before proceeding to the one-loop expression, it will be useful to have the tree-level expression close to the alignment limit. Using $\sin(\beta-\alpha)=\sin(\pi/2+x)=\cos(x)\sim 1-x^{2}/2$, we obtain
\begin{align}\label{eq:kanemura_tree_level}
    \kappa_{\lambda}^{(0)} = &1+\frac{3}{2}x^2-2x^3\cot(2\beta)-\frac{13x^4}{8}-\left(2x^2-2x^3\cot(2\beta)-\frac{5x^4}{3}\right)\frac{M^{2}}{m_{h}^{2}}
\end{align}

We notice that from $\mathcal{O}(x^{3})$, there is a dependence on $\tan\beta$. At loop order, considering terms up to $\mathcal{O}(x^{2})$ as well as loops containing either BSM particles or the top, we obtain~\cite{Kanemura:2004mg} 
\begin{align}\label{eq:hhh}
    \kappa_{\lambda}^{\mbox{\tiny{2HDM}}} = & 1  + \frac{3}{2} 
                   \left(1 - \frac{4 M^2}{3 m_h^2}\right) x^2
              + \frac{m_{H}^4}{12 \pi^2 m_h^2 v^2} 
                         \left(1 - \frac{M^2}{m_H^2}\right)^3 
              + \frac{m_{A}^4}{12 \pi^2 m_h^2 v^2} 
                         \left(1 - \frac{M^2}{m_A^2}\right)^3 \nonumber\\
  &
              + \frac{m_{H^\pm}^4}{6 \pi^2 m_h^2 v^2} 
                         \left(1 - \frac{M^2}{m_{H^\pm}^2}\right)^3
              - \frac{ m_t^4}{ \pi^2 m_h^2 v^2} 
\end{align}

Notice that the top contribution decreases $\kappa_{\lambda}$. It is also clear that $\kappa_{\lambda}$ deviates from unity for large splitting between $M$ and the BSM scalars. In the 2HDM in general, $M$ can be chosen as a free parameter, so one has the freedom to maximize $\kappa_{\lambda}$ by a suitable adjustment of $M$ (in particular, by adopting a large splitting between $M$ and the BSM scalar masses). In the 3-3-1 EFT it is no longer allowed, since $M$ is given by 
\begin{align}
M^{2}&=m_{A}^{2}+v^{2}\left(\Lambda_{5}+\frac{1}{2}\Lambda_{6}t_{\beta}^{-1}+\frac{1}{2}\Lambda_{7}t_{\beta}\right)\nonumber\\
&\sim m_{A}^{2}-\left(\frac{f^{2}}{m^{2}_{\chi_{0}}}\right)v^{2}+\frac{1}{2}\left(\frac{2\sqrt{2}fv_{\chi}\lambda_{13}}{m^{2}_{\chi_{0}}}\right)t_{\beta}^{-1}+    \frac{1}{2}\left(\frac{2\sqrt{2}fv_{\chi}\lambda_{23}}{m^{2}_{\chi_{0}}}\right)t_{\beta}   
\end{align}
where $m_{A}^{2}=\frac{fv_{\chi}}{c_{\beta}s_{\beta}}+\frac{fv^{2}c_{\beta}s_{\beta}}{v_{\chi}}\sim\frac{fv_{\chi}}{c_{\beta}s_{\beta}}$, and for simplicity, we are not showing the one-loop matching terms. We are interested in the regime where $m_{A}$ is relatively light (up to TeV), while $v_{\chi}\gg v$. Therefore, we are required to choose $f \ll v_{\chi}$. Since the 3-3-1 EFT is defined in the regime that $m_{\chi_0}\sim v_{\chi}$, we obtain that $M^{2}\sim m_{A}^{2}$, implying
\begin{align}\label{eq:331_leading}
    \kappa_{\lambda}^{331} \sim 
      & 1  + \frac{3}{2} 
                   \left(1 - \frac{4 m_{A}^2}{3 m_h^2}\right) x^2
             - \frac{ m_t^4}{ \pi^2 m_h^2 v^2} + \frac{m_{H}^4}{12 \pi^2 m_h^2 v^2} 
                         \left(1 - \frac{m_{A}^2}{m_H^2}\right)^3 
             \nonumber\\
  &
              + \frac{m_{H^\pm}^4}{6 \pi^2 m_h^2 v^2} 
                         \left(1 - \frac{m_{A}^2}{m_{H^\pm}^2}\right)^3
   \end{align}

The formula above represents the leading approximation for $\kappa_{\lambda}^{331}$, where the terms due to the heavy spectrum are not shown. We will discuss further in the section how accurate this approximation is in view of the present phenomenological constraints. However, there are other aspects that \cref{eq:hhh} fails to uncover, even in the 2HDM. Therefore, we proceed to include some refinements. First, in the 2HDM, the normalized top Yukawa is given by
\begin{equation}\label{eq:top_alig}
    \xi_u = \left(\frac{\cos{(\beta-\alpha})}{\tan{\beta}}+\sin{(\beta-\alpha})\right)    
\end{equation}

Also, following \cite{Senaha:2018xek}, in the SM case there are sub-leading contributions from the top, as well as loop corrections containing the SM Higgs
\begin{align}\label{eq:SM}
\lambda_{hhh}^{sub} = &- \frac{3 m_h^2}{v}
      \left\{ \frac{m_{h}^2}{2 \pi^2 v^2}+ \frac{7 m_t^2}{32 \pi^2  v^2}\right\} 
\end{align}

Moreover, as we deviate from the alignment limit, there are corrections from the other trilinear scalar couplings, which appear in triangle diagrams containing BSM scalars. Since these couplings are behind the appearance of the factors $\left(1-\frac{M^{2}}{m^{2}_{\text{BSM}}}\right)^3$ in~\cref{eq:hhh}, we can just adapt them from the general non-alignment case. By inspection of the scalar couplings in the 2HDM, we will perform the replacements
\begin{align}
    \left(1-\frac{M^{2}}{m^{2}_{A}}\right)&\rightarrow  \left(1-\frac{M^{2}}{m^{2}_{A}}\right) + \frac{M^{2}-m_{h}^{2}}{2m_{A}^{2}}\left(t_\beta-\frac{1}{t_\beta}\right)x\\
    \left(1-\frac{M^{2}}{m^{2}_{H^{\pm}}}\right)&\rightarrow  \left(1-\frac{M^{2}}{m^{2}_{H^{\pm}}}\right) + \frac{M^{2}-m_{h}^{2}}{2m_{H^{\pm}}^{2}}\left(t_\beta-\frac{1}{t_\beta}\right)x\\
    \left(1-\frac{M^{2}}{m^{2}_{H}}\right)&\rightarrow  \left(1-\frac{M^{2}}{m^{2}_{H}}\right) - \frac{2 m_{H}^{2}+m_{h}^{2}-3M^{2}}{2m_{H}^{2}}\left(t_\beta-\frac{1}{t_\beta}\right)x
\end{align}

Finally, $\kappa_{\lambda}^{331}$, including the subleading corrections and keeping terms up to $\mathcal{O}(x^{3})$ will be given by (once again, we do not show terms depending on the heavy spectrum for simplicity. They can be recovered in \cref{ap:one-loop})
\begin{align}\label{eq:331}
    \kappa_{\lambda}^{331} \sim  
      &1  + \frac{3}{2} 
                   \left(1 - \frac{4 m_{A}^2}{3 m_h^2}\right) x^2+\frac{m_{A}^{2}-m_{h}^{2}}{m_{h}^{2}}\left(t_\beta-\frac{1}{t_\beta}\right)x^{3}
             - \frac{ m_t^4}{ \pi^2 m_h^2 v^2}\left(1-\frac{7m_{h}^{2}}{32m_{t}^{2}}\right)\xi_{u}^{3} 
             \nonumber\\
  &
+\frac{m_{h}^2}{2 \pi^2 v^2}+ \frac{m_{H}^4}{12 \pi^2 m_h^2 v^2}                          \left[1-\frac{m_{A}^{2}}{m^{2}_{H}} + \frac{2 m_{H}^{2}+m_{h}^{2}-3m_{A}^{2}}{2m_{H}^{2}}\left(t_\beta-\frac{1}{t_\beta}\right)x\right]^{3}\nonumber\\
  &
  + \frac{m_{H^\pm}^4}{6 \pi^2 m_h^2 v^2}                          \left[1-\frac{m_{A}^{2}}{m^{2}_{H^{\pm}}} + \frac{m_{A}^{2}-m_{h}^{2}}{2m_{H^{\pm}}^{2}}\left(t_\beta-\frac{1}{t_\beta}\right)x\right]^{3}
      + \frac{(m_{A}^{2}-m_{h}^{2})^{3}x^{3}}{96 \pi^2 m_h^2 m_{A}^{2} v^2}                          \left(t_\beta-\frac{1}{t_\beta}\right)^{3}
\end{align}

\begin{figure}[ht!]
        \centering
         \includegraphics{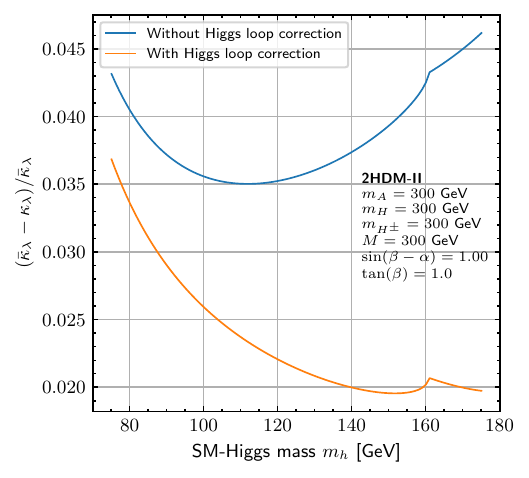}
        \caption{Deviation from the \textit{anyH3} prediction, with and without including the Higgs loop, see~\cref{eq:SM}. We choose the benchmark values shown in the figure, rendering contributions only from SM particles.}
        \label{fig:kappa_mh}
    \end{figure}

In the remainder of the section, we discuss the agreement between the last formula (for 3-3-1 EFT as well as general 2HDM) and the full calculation obtained by employing \textit{anyH3}~\cite{Bahl:2023eau}. We will first consider the case where the terms related to the heavy spectrum are neglected. We will discuss their importance at the end of the section. We include in the legends of the following figures if we are considering the 2HDM in general (for which $M$ is a free parameter) or the 3-3-1 EFT. In~\cref{fig:kappa_mh} we show the difference between the approximate ($\bar{\kappa}_{\lambda}$) and the full formula ($\kappa_{\lambda}$), in terms of the SM Higgs mass. Since we consider the alignment limit ($s_{\beta-\alpha}=1$), and degenerate masses, we recover the SM result. Notice the importance of the subleading correction due to the SM Higgs loop given by~\cref{eq:SM} (orange line) for a better agreement, in particular for higher masses. 

\begin{figure}[ht!]
        \centering
        \includegraphics{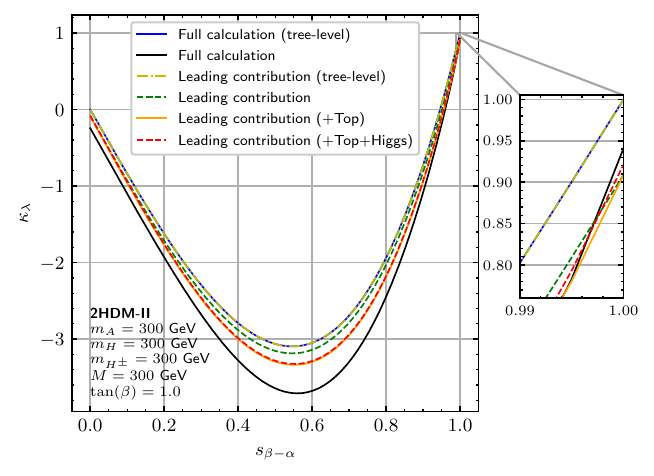}
        \caption{Dependence of $\kappa_{\lambda}$ on $s_{\beta-\alpha}$ for the 2HDM type II. We choose the benchmark values showed in the figure. The black line is for the full calculation using \textit{anyH3}, while the blue line is considering only the tree-level contribution. The light (dark) green line is considering tree (one-loop) contribution from~\cref{eq:hhh}, while the orange and red lines are including the top (subleading) and higgs loop corrections given by~\cref{eq:SM}.}
        \label{fig:kappa_sba}
    \end{figure}

Next, we discuss the dependence of $\kappa_{\lambda}$ on $s_{\beta-\alpha}$. We recall that $s_{\beta-\alpha}=1$ implies that the lightest of the CP-even scalars in the 2HDM behaves precisely as the SM Higgs. In order to remove the dependence of the BSM scalar loops, we choose as benchmark a degenerate scenario. As~\cref{fig:kappa_sba} shows, the inclusion of one-loop corrections will decrease the value of $\kappa_{\lambda}$, and the sub-leading corrections are important (in particular related to the top loop) to obtain a better agreement to the full calculation. The difference is more pronounced for values of $s_{\beta-\alpha}\sim 0.5$, which are disfavored by present data from colliders \cite{ATLAS:2024lyh}. In particular, the bound $s_{\beta-\alpha}>0.995$ is enforced for type II 2HDM at $95\%$ C.L. For this range, the leading contribution (including the top and Higgs corrections) is at most $2\%$ lower than the full calculation.  

\begin{figure}[h!]
    \centering
\includegraphics[scale=1]{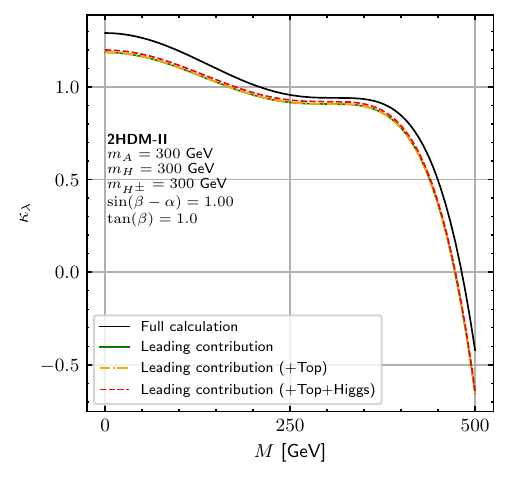}
    \caption{Dependence of  $\kappa_{\lambda}$ on M for the 2HDM type II. We choose the benchmark values showed in the figure. The black line is for the calculation using \textit{anyH3}, the green line is considering~\cref{eq:hhh}, while the orange and red lines are including the top (subleading) and Higgs loop corrections given by~\cref{eq:SM}.}
    \label{fig:kappa_M}
\end{figure}

\begin{figure}[h!]
    \centering
    \includegraphics[width=0.45\linewidth]{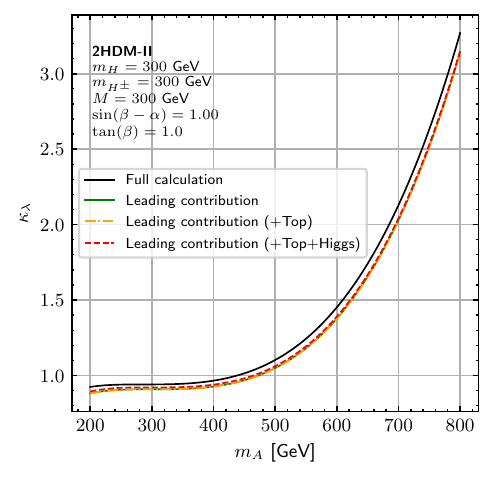}
    \includegraphics[width=0.45\linewidth]{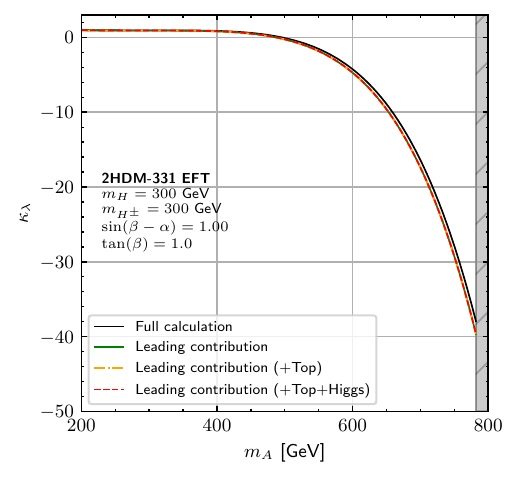}
    \caption{Dependence of  $\kappa_{\lambda}$ on $M_{A}$ for the 2HDM type II (left) and the 3-3-1 EFT (right). We choose the benchmark values showed in the figure. The black line is for the calculation using \textit{anyH3}, the green line is considering~\cref{eq:331_leading}, while the orange and red lines are including the top (subleading) and Higgs loop corrections given by~\cref{eq:SM}. The gray area shows the region excluded by the perturbative unitarity constraint.}
    \label{fig:331mAalign}
\end{figure}

Next, we discuss the influence of $M$ on $\kappa_{\lambda}$, as shown in~\cref{fig:kappa_M}. For simplicity, we consider the alignment scenario. At $M=300~\mbox{GeV}$ we recover the SM, for which $\kappa_{\lambda}<1$ (recall that the top contribution is negative). As $M$ increases, all corrections containing the BSM scalars also become negative, decreasing the value of $\kappa_{\lambda}$. We can, nevertheless, single out only one of the BSM scalars contributions, by choosing degenerate masses and a suitable $M$ as done in the left plot of~\cref{fig:331mAalign}. Notice that as $m_{A}$ increases with \textbf{fixed} $M$, the value of $\kappa_{\lambda}$ also increases, behaving roughly as $\kappa_{\lambda} \sim m_{A}^{4}(1-3M^{2}/m_{A}^2)$. On the other hand, for the 3-3-1 EFT, $M$ is no longer a free parameter, implying that we cannot remove the dependence on the other BSM scalar loops. Actually, the contribution from the loop containing the pseudoscalar $A$ vanishes, see~\cref{eq:331}. Therefore, as $m_{A}$ increases, as shown in the right plot of~\cref{fig:331mAalign}, $\kappa_{\lambda}$ attains negative values.

\begin{figure}[h!]
        \centering
        \includegraphics[scale=1]{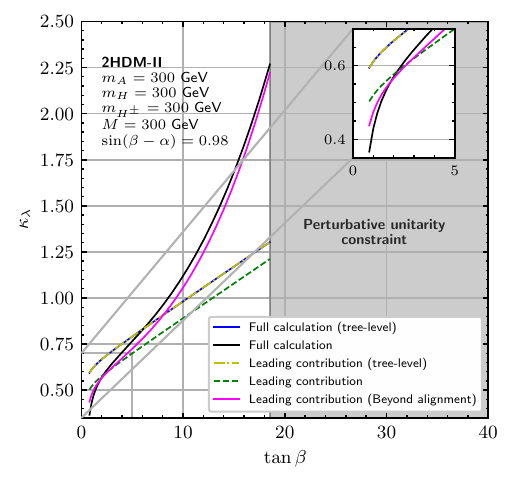}
        \caption{Dependence of  $\kappa_{\lambda}$ on $\tan\beta$  for the 2HDM type II. We choose the benchmark values showed in the figure. The black line is for the full calculation using \textit{anyH3}, while the blue line is considering only the tree-level contribution. The light (dark) green line is considering tree (full) contribution from~\cref{eq:331_leading}, while the pink line is including all refinements given in~\cref{eq:331}. The gray area shows the region excluded by the perturbative unitarity constraint.}
        \label{fig:tanbeta}
    \end{figure}

Next, we discuss the influence of $\tan\beta$ on $\kappa_{\lambda}$ in~\cref{fig:tanbeta}. Since the top Yukawa is already close to the perturbativity limit, the region with $t_{\beta}<1$ is disfavored. Therefore, we will focus on $t_{\beta}>1$. For this range, the influence of $\zeta_{u}$ is marginal, while the terms proportional to $t_\beta$ will be the leading ones, in particular for large values of this parameter. Notice also that according to~\cref{eq:331}, the dependence on $t_{\beta}$ only appears if one allows deviations from the alignment limit. For degenerate masses, in particular, only corrections proportional to $x^{3}$ survive. Once these terms are included, we find a good agreement with the complete calculation (see the pink curve of~\cref{fig:tanbeta}.)

\begin{figure}[h!]
    \centering
    \includegraphics[scale=0.8]{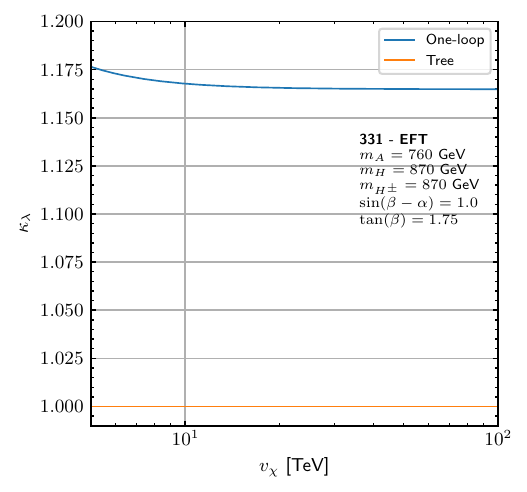}
    \caption{Trilinear Higgs coupling normalized by the SM tree-level value with the full matching to the 331 - EFT as function of the EFT scale, $v_\chi$.}
    \label{fig:kappa_final-331}
\end{figure}

Finally, we discuss the influence of the heavier spectrum on $\kappa_{\lambda}^{331}$. Their impact comes both from the inclusion of dimension-six terms as well as from considering one-loop matching for the quartic couplings. The results are lengthy, so we refrain from presenting them here. They can be found in the auxiliary notebook\footnote{\url{https://github.com/LeoFerreira8/hhh---General}}. In order to gain some insight about their influence, we chose the value of the heavy scale to be $v_{\chi}=10\;\rm{TeV}$, which is the lowest value allowed by direct searches~\cite{Alves:2022hcp}. For simplicity, all the heavy spectrum is considered to be degenerate, with masses also at 10 TeV. In \cref{fig:kappa_final-331} we show how the tree-level and one-loop contribution to $\kappa_{\lambda}^{331}$ varies in terms of the heavy scale. We have chosen a benchmark that maximizes the ratio between $\kappa_{\lambda}^{331}$ and $\kappa_{\lambda}^{\rm{2HDM}}$ with $\Lambda_{5,6,7}=0$. 
As can be seen in the plot, the deviation from the limit $v_{\chi}\rightarrow\infty$ to the case for $v_{\chi}=10\;\rm{TeV}$ amounts to few percent. In the next section, we will perform a broader numerical scan, showing that this behavior is still maintained.

In summary, larger values for $\kappa_{\lambda}$ are attained in the 3-3-1 EFT mainly by allowing a large split between the pseudo-scalar mass and the other BSM scalar masses. Large values for $\tan\beta$ can also play an important role, when one departs from the alignment limit. Given present constraints, including the heavier spectrum contribution is important for a precise determination of the trilinear Higgs coupling, however, their correction is at the percent level.

We conclude this section briefly discussing the importance of two-loop corrections for $\kappa_{\lambda}$. As shown in~\cite{Bahl:2022jnx}, these can be relevant and even surpass the one-loop corrections. However, they are only known in the alignment regime~\cite{Braathen:2019pxr,Braathen:2019zoh}. Since we aim to perform a comprehensive study of how large $\kappa_{\lambda}$ can be in the 3-3-1 EFT, we opt to consider a non-alignment scenario in general, for which the two-loop corrections are not known. Thus, we restrain our analysis for $\kappa_{\lambda}$ at the one-loop level. Nevertheless, in \cref{ap:two-loop} we discuss how our findings for an alignment scenario would be affected by the inclusion of two-loop corrections.

\section{Phenomenology}
\label{sec:pheno}

Once we have established the most promising conditions to ensure that a large $\kappa_{\lambda}$ possibly detectable at the high luminosity regime of the LHC, we
proceed to impose constraints from distinct sources on our model, including theoretical, flavor, and collider. We will describe each of them in the following subsections. In order to assess the influence of those constraints in the multi-dimensional parameter space of our model, we will perform a scan on the mixing angle $s_{\beta - \alpha}$, the physical masses $m_{A}$, $m_{H}$, $m_{H^{\pm}}$, $t_{\beta}$, and $M$. For the 2HDM, we will vary $M$ freely. For the 3-3-1 EFT we will consider two cases. In the first, we impose that $\Lambda_{5}<10^{-8}$, which stands for $v_{\chi}\sim 100\;\rm{TeV}$ (the limit to the right of \cref{fig:kappa_final-331}). In the second, we consider $v_{\chi}\sim 10\;\rm{TeV}$. This amounts to non-null values for $\Lambda_{5,6,7}$. For simplicity, we will discuss the latter only in the summary subsection, after all constraints are imposed. The calculation of $\kappa_{\lambda}$ will be performed by employing \textit{anyBSM}~\cite{Bahl:2023eau}. For the 2HDM scenario, we will use an UFO file derived for a Type II case with $\Lambda_{6,7}=0$. For the 3-3-1 scenario, we have created a UFO file where $\Lambda_{5,6,7}$ are arbitrary, which we choose to match the formulas derived in \cref{ap:one-loop}. Moreover, we also include dimension-six terms to the result for $\kappa_{\lambda}$ obtained from \textit{anyBSM}. We will adopt the following ranges as shown in~\cref{tab:range}.
\begin{table}[h!]
    \centering
    \begin{tabular}{|c|c|c|c|c|c|c|c|}
    \hline
     Parameter & $s_{\beta - \alpha}$ & $m_{A}$ (GeV) & $m_{H}$ (GeV) & $m_{H^{\pm}}$ (GeV) &  $t_{\beta}$ & $M$ $(\mbox{GeV})$ \\
     \hline
     Range & (0.98 - 1) & (125 - 1500) & (125 - 1500) & (125 - 1500) & (0.8 - 40) & (100 - 5000)  \\
     \hline
    \end{tabular} 
    \caption{Range of the free parameters used on the scan. When considering alignment, we do not allow $s_{\beta - \alpha}$ to vary freely, and when considering the 3-3-1 EFT, we do not vary M freely.}
    \label{tab:range}
\end{table}

 We scan the parameter space until around three to five thousand points pass all the imposed constraints. Then, after establishing the regions where the coupling is larger, we perform a dedicated scan up to, again, three to five thousand points in those regions to find the maximal value. Thus, we consider, in total, around 10,000 points of data in our scan that comply with every bound considered.

\subsection{Theoretical}

As customary in models with multiple scalars, we have to enforce the stability of scalar potential (bounded from below conditions), pertubative of its coupling as well as pertubative unitarity for the scattering matrix. As shown in~\cite{Cherchiglia:2022zfy}, all these theoretical conditions, for the 3-3-1 EFT, can be extracted from the 2HDM~\cite{Maniatis:2006fs,Ginzburg:2005dt}. We have employed our own implementation for stability and pertubative conditions. For pertubative unitarity (PU) we relied on the implementation available at \textit{anyBSM}~\cite{Bahl:2023eau}, which computes the eigenvalue $a_{0}$ for the $SS\rightarrow SS$ scattering matrix for each of the scalars $S=A, H, H^{\pm}$ and automatically selects the maximum value. Since the PU bound will play a decisive role in constraining the parameter space of our model, we opted to show it as a separate constraint in our scan and consider $|{\rm{Re}}(a_0)|<0.5$. For the other two constraints, we will reserve the letters P for perturbativity of the quartic couplings (assumed to be smaller than $4\pi$) and S for the stability of the scalar potential. 

\begin{figure}[h!]
    \centering
    \includegraphics[scale=1]{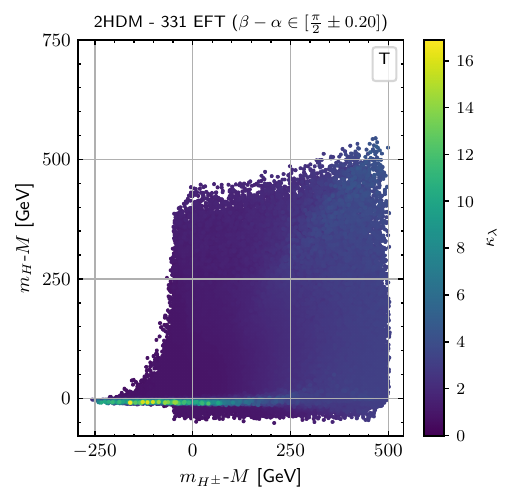}
    \caption{Parameter space of mass difference for points that comply with all theoretical bounds (perturbativity (P), stability (S) and unitarity (PU) = (T)) as function of the value of $\kappa_\lambda$.}
    \label{fig:Teo_kappa}
\end{figure}

\begin{figure}[h!]
    \centering
    \includegraphics[width=0.45\textwidth]{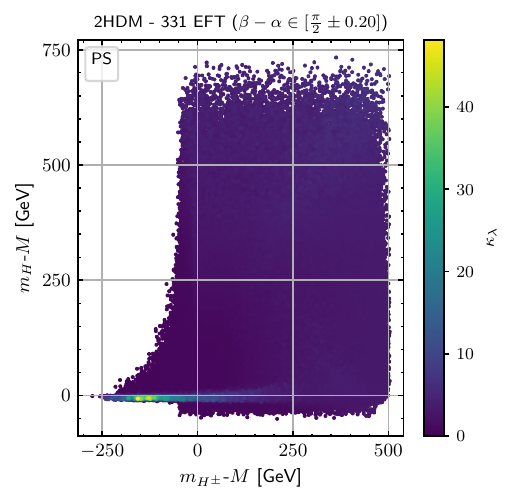}
    \includegraphics[width=0.45\textwidth]{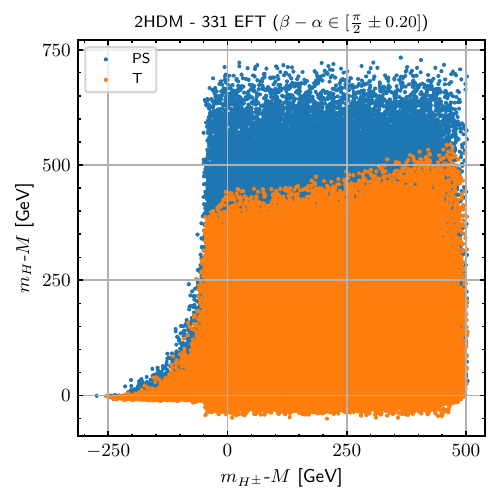}
    \caption{Parameter space of mass difference for points that comply with pertubativity (P) and vacuum stability (S) bounds as function of the value of $\kappa_\lambda$ (left) and the comparison of the former case with the parameter space allowed region for all of the theoretical bounds (T), that is, including the perturbative unitarity bounds (PU) (right).}
    \label{fig:Teo_kappa_comp}
\end{figure}

We show in~\cref{fig:Teo_kappa} how the theoretical bounds limit the parameter space of the mass differences and the maximal value of $\kappa_\lambda$. Recall that, in the 3-3-1 EFT, $M\sim m_{A}$.
Higher values for $\kappa_\lambda$ are obtained in the regions $m_{H}^{\pm}<m_{A}$ and $m_{H}\sim m_{A}$. As mentioned, the perturbative unitarity bound plays a critical role in constraining the parameter space for critical values of $\kappa_\lambda$. In~\cref{fig:Teo_kappa_comp} (left) we show that PU forbids a large splitting between $m_{H}$ and $m_{A}$. Moreover, it will be crucial to remove the points with large $t_{\beta}$, as presented in~\cref{fig:tanbeta}. In~\cref{fig:Teo_kappa_comp} (right) we show how~\cref{fig:Teo_kappa} would appear if PU were not enforced. It is possible to see that the perturbative unitarity bound restricts the maximal value of the trilinear Higgs coupling to less than half of the maximum allowed value from the other theoretical constraints, due to the removal of regions with large $\tan{\beta}$.

\subsection{Electroweak constraints}

Next, we consider the S, T, and U parameters. In 2022, the CDF collaboration released an updated value for the W boson mass, which differs by more than $7\sigma$ from the SM prediction~\cite{CDF:2022hxs}. More recently, the ATLAS and CMS collaborations provided an updated analysis, which did not find a significant deviation from previous measurements~\cite{ATLAS:2024erm,CMS:2024nau}. Until this tension is settled, we opt to consider two cases in the present work. In the first, we assume the values for the S, T parameters (hereafter we assume $\Delta U=0$) obtained by employing the W mass reported from the ATLAS collaboration~\cite{ParticleDataGroup:2024cfk}
\begin{equation}\label{eq:WPDG}
    \Delta S = -0.05 \pm 0.07, \quad \Delta T = 0.00 \pm 0.06, \quad \text{correlation = } 0.93
\end{equation}
The second case is by considering only the CDF measurement for the W mass~\cite{Lu:2022bgw} 
\begin{equation}\label{eq:WCDF}
    \Delta S = 0.15 \pm 0.08, \quad \Delta T = 0.27 \pm 0.06, \quad \text{correlation = } 0.93
\end{equation}

For the evaluation of the S, T parameters, we employed \textit{SPheno}~\cite{Porod:2003um,Porod:2011nf} within the 2HDM. For the 3-3-1 EFT, we further consider that $\Lambda_{5}\sim 0$. Since these parameters are obtained from radiative corrections, the heavy spectrum of the 3-3-1 model could play a role. We have explicitly checked that this is not the case, employing the formulas for the S, T parameter in the 3-3-1 model in general~\cite{Liu:1993fwa}.

\begin{figure}[h!]
    \centering
    \includegraphics[scale=1]{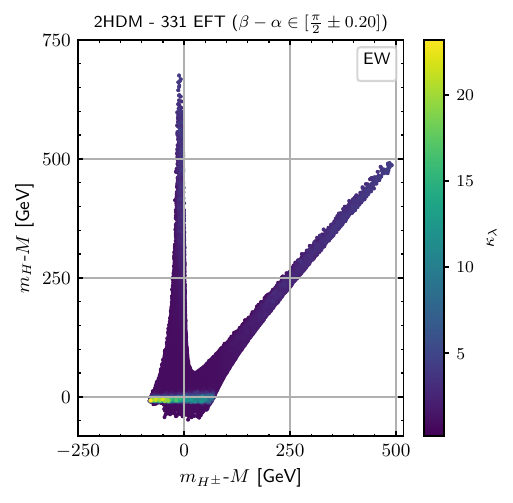}
    \caption{Parameter space of mass difference for points that comply with electroweak precision observable bounds (from PDG) as function of the value of $\kappa_\lambda$.}
    \label{fig:stu_kappa}
\end{figure}

\begin{figure}[h!]
    \centering
    \includegraphics[width=0.45\textwidth]{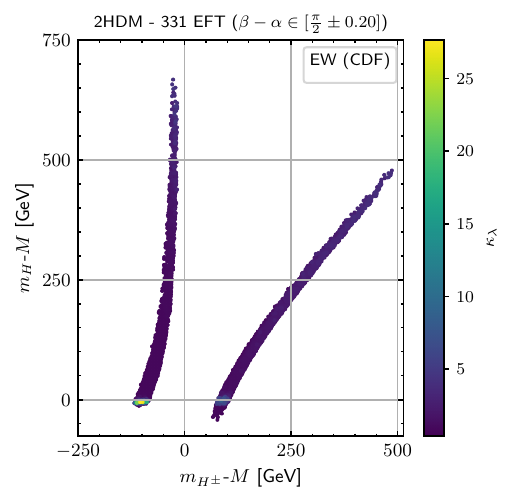}
    \includegraphics[width=0.45\textwidth]{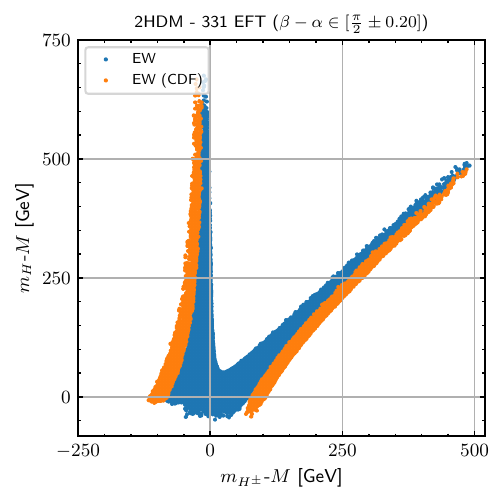}
    \caption{Parameter space of mass difference for points that comply with electroweak precision observable bounds (from CDF measurement) as function of the value of $\kappa_\lambda$ (left) and the comparison of the former case with the parameter space allowed for the PDG global fit (as shown in~\cref{fig:stu_kappa}) (right). }
    \label{fig:stu_kappa_comp}
\end{figure}

In~\cref{fig:stu_kappa}, we show how the electroweak precision bounds (EW) constrain our main parameter space. All points comply with PS (from theoretical bounds).
We note that EW is one of the main constraints for the parameter space and also helps to reduce the maximal value of $\kappa_\lambda$, although it is not the main factor (compared to the right figure in~\cref{fig:Teo_kappa_comp}). In particular, it enforces either $m_{H^{\pm}}\sim m_{A}$ or $m_{H^{\pm}}\sim m_{H}$. In~\cref{fig:stu_kappa_comp}, we show how the measurement from CDF for the $W$ mass changes the allowed parameter space region. We see that the maximum value of $\kappa_\lambda$ in this case is slightly higher than the PDG value. We also show the comparison between the parameter space in each case and note that the regions are almost disjoint, as would be expected since the two results are in tension with one another. 

\subsection{Collider}
In order to systematically incorporate the bounds coming from colliders into our numerical scan, we make use of \textit{HiggsTools}~\cite{Bahl:2022igd,Bechtle:2013wla,Bechtle:2020pkv,Bechtle:2013xfa,Bechtle:2020uwn}, a successor to the two packages \textit{HiggsSignals} and \textit{HiggsBounds}. \textit{HiggsSignals} is used to reinterpret the known 125 GeV-Higgs resonances found in multiple channels, as well as its properties, in the case of extra scalar fields. Employing a $\chi^2$ distribution, we quantify the agreement between the model predictions for given parameters with the available data, excluding points that do not fit the known resonances at the $95\%$ confidence level. With \textit{HiggsBounds}, we also employ a $\chi^2$ distribution to quantify the exclusion of given model parameters, based on direct searches in the LHC for scalar particle signals. \textit{Higgsbounds} automatically selects the channel that provides the most stringent constraints, evaluating whether such a set of parameters is excluded or not at the $95\%$ confidence level. We find that the channels that are most constraining for the parameter space are $h, H\rightarrow{} ZZ \rightarrow{} 4\ell,\,2\ell2\nu$; $h\rightarrow{} \gamma\gamma, ZZ, WW,\tau\tau, bb$~\cite{CMS:2013fjq,CMS-PAS-HIG-12-045,ATLAS:2020tlo}; $gg\rightarrow{} A \rightarrow{} Z \,h \, (\bar{b}b)$~\cite{CMS:2019qcx}; $gg\rightarrow{} H,A \rightarrow{} \tau\tau \,(\bar{b}b)$~\cite{CMS:2022goy}; $A\rightarrow{} Z\,H \rightarrow{} \ell \ell\, W W, \ell \ell \, b b$~\cite{ATLAS:2020gxx}; and $t\rightarrow{} b \,H^\pm \rightarrow{} b\, \tau^\pm \, \nu$~\cite{ATLAS:2018gfm}, but other channels were also considered.

Among the previously considered channels, the scalar-vector and scalar-scalar couplings are mostly sensitive to deviations from the alignment situation. Specifically, the decay rates $h\rightarrow{} ZZ, \gamma\gamma, WW \propto s_{\beta\alpha}$, while $H\rightarrow{} ZZ, \gamma\gamma, WW$ and $A\rightarrow{}Z\,H$ are proportional to $c_{\beta-\alpha}$. Thus, when the alignment condition is met, we recover the SM case and the bounds are weaker. The scalar-fermionic couplings, on the other hand, are mostly proportional to $t_{\beta}$, so $h,H,A\rightarrow{}\tau\tau, bb$ give us the strongest constraints on $\tan{\beta}$\footnote{This is true mostly for Type II 2HDM, where the couplings to down-type quarks and leptons are proportional to $\tan{\beta}$~\cite{Branco:2011iw}. For Type I, it is still possible to impose such constraints, but they will be significantly weaker.}.

\begin{figure}[h!]
    \centering
    \includegraphics[scale=1]{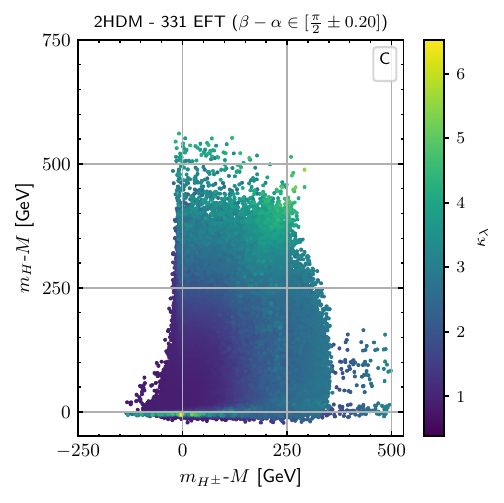}
    \caption{Parameter space of mass difference for points that comply with collider bounds as function of the value of $\kappa_\lambda$.}
    \label{fig:collider_kappa}
\end{figure}

\begin{figure}[h!]
    \centering
    \includegraphics[width=0.45\textwidth]{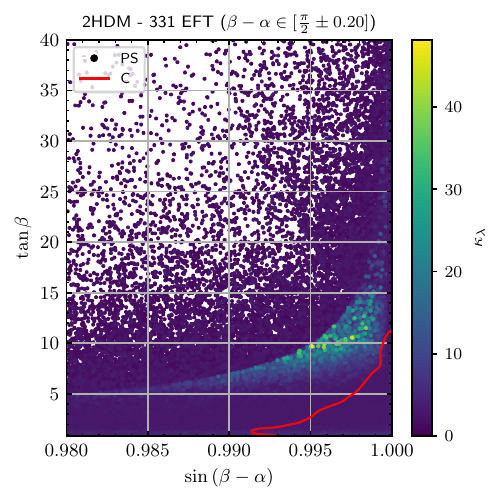}
    \includegraphics[width=0.45\textwidth]{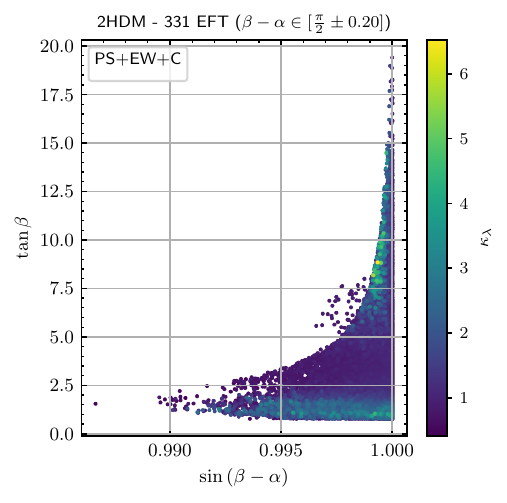}
    \caption{Parameter space of $\sin{(\beta-\alpha)}$ by $\tan{\beta}$ as a function of the value of $\kappa_\lambda$. On the left, we show all the points that pass the perturbativity and stability tests compared to the approximate region of the points that pass collider contraints (enclosed by the red curve). On the right, we show the points that pass all the bounds discussed so far.}
    \label{fig:collider_kappa_comp}
\end{figure}

In~\cref{fig:collider_kappa} we show the influence of collider bounds on the parameter space of our model (to be compared with the right~\cref{fig:Teo_kappa_comp}). 
Notice that the maximal value for $\kappa_{\lambda}$ is drastically reduced, although a relatively large splitting between the scalars is still allowed. The main reason behind this behavior is the constraint on the deviation from the alignment limit ($s_{\beta-\alpha}$) as well as the maximal value for $t_{\beta}$, as can be seen in~\cref{fig:collider_kappa_comp}. On the left, we show every point in this parameter space that is allowed by perturbativity and vacuum stability, with its respective value of $\kappa_\lambda$. In the same plot, we show in red the approximate region containing the majority of points allowed by collider constraints. On the right, we show how all the bounds discussed so far constrain this parameter space and the maximal value of $\kappa_\lambda$.
As we showed before, $\kappa_{\lambda}$ increases with $t_{\beta}$, even in a mass-degenerate scenario. However, this behavior is only possible if deviations from the alignment limit are enforced. In~\cref{fig:collider_kappa_comp} one can see that the collider bounds are very restrictive to $s_{\beta-\alpha}$, constraining it to the interval $\left[0.990,1\right]$, as well as limiting the maximal value of $\tan{\beta}$, which is restricted to $\tan{\beta}<20$.

\subsection{$B\rightarrow X_{s} \gamma$}

\begin{figure}[h!]
    \centering
    \includegraphics[scale=1]{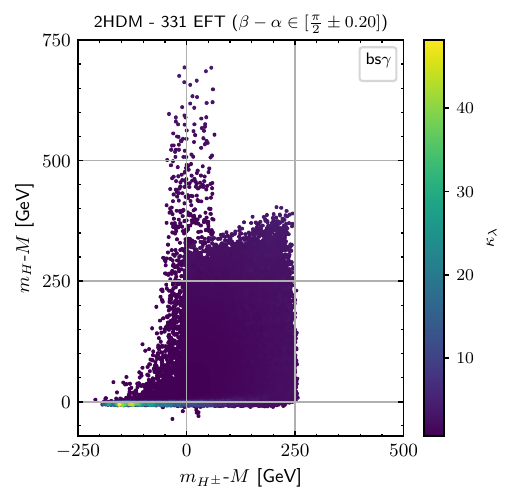}
    \caption{Parameter space of mass difference for points that comply with $b\rightarrow s \gamma$ present bounds, as a function of the value of $\kappa_\lambda$.}
    \label{fig:bsg_kappa}
\end{figure}

An stringent constraint for 2HDM, in general, is due to the decay $B\rightarrow X_{s} \gamma$, which is well in agreement with the prediction of the SM. This bound is particularly important for Type II, since there is an enhancement of the coupling between the charged scalar and the bottom quark given by $t_{\beta}$ in this case. Using the analytical formulas of~\cite{Enomoto:2015wbn}, one finds a lower bound on the charged scalar mass of around 600 GeV, which is independent of the value of $t_{\beta}$. For the 3-3-1 EFT, there is more freedom when defining the Yukawa sector. In particular, this model allows for the physical realization of the mixing matrices $V_{u}$ and $V_{d}$, as opposed to the SM (or 2HDM) where only their product $V_{u}^{\dagger}V_{d}$ is physical (parameterized as the CKM matrix). By adopting specific choices for $V_{u}$, it is possible to relax the constraint on the charged scalar mass~\cite{Fan:2022dye,Cherchiglia:2022zfy,Doff:2024hid}. However, in our work, in order to be conservative, we will adopt $V_{u}$ as the identity matrix, which renders $V_{d}$ as the CKM matrix. It is straightforward to implement the results of Ref.~\cite{Enomoto:2015wbn} for our model, rendering the allowed parameter space depicted in ~\cref{fig:bsg_kappa}. 

\subsection{Maximal $\lambda_{hhh}$ in the 3-3-1 EFT}

\begin{figure}[h!]
    \centering
    \includegraphics[width=0.45\textwidth]{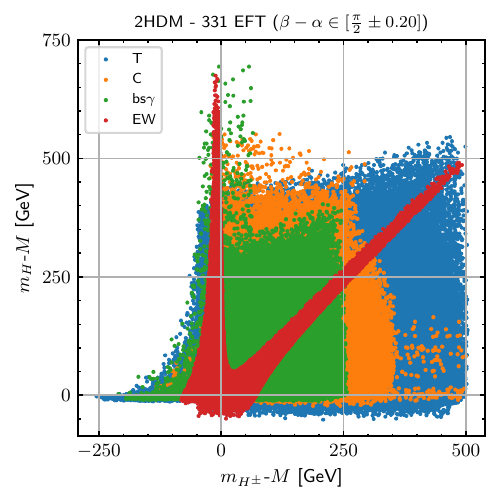}
    \includegraphics[width=0.45\textwidth]{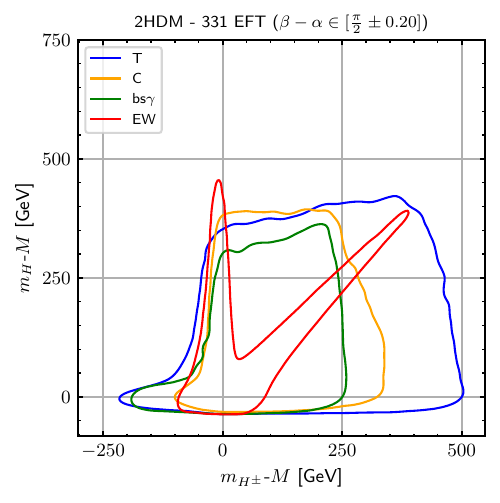}
    \caption{Comparison of points that comply with each set of bounds, theoretical (T), collider (C), flavor ($b s\gamma$), and electroweak precision observables (EW).}
    \label{fig:Comparison_final}
\end{figure}

To summarize, we see that by imposing only the theoretical bounds (PS), $\kappa_{\lambda}$ can range up to 45; see~\cref{fig:Teo_kappa_comp}. When constraints from flavor ($b s\gamma$) are also included, we still attain the same large value for $\kappa_{\lambda}$. However, the available parameter space is significantly reduced; see~\cref{fig:bsg_kappa}. More importantly is the addition of electroweak precision observables (EW). It reduces $\kappa_{\lambda}$, which can be up to 25, as well as the available parameter space; see~\cref{fig:stu_kappa}. Nevertheless, the main constraint is provided by collider searches (C), which renders $\kappa_{\lambda}$ to be of order 6, see~\cref{fig:collider_kappa}. In order to highlight the importance of the distinct constraints in restricting the parameter space, we show in~\cref{fig:Comparison_final} every set of bounds superposed.

The theoretical bounds (T), shown in blue, restrict $m_{H^\pm}$ to be at most 500 GeV higher than $m_{A}$, while the splitting between $m_{H}$ and $m_{A}$ is generally a little smaller. Moreover, the region where ${A}$ is the heaviest of the BSM scalars is significantly constrained. However, these findings do not restrict the maximal value of $\kappa_\lambda$, which can still be as high as 17. 
For the electroweak precision observables (EW), in red, we notice that they are important for restricting the splitting between the BSM scalar masses. In particular, it is not possible to have simultaneously $m_{H}$ and $m_{A}$ far from $m_{H^{\pm}}$. However, in respect of $\kappa_\lambda$, these constraints are weaker than T, as can be seen in~\cref{fig:stu_kappa}, where $\kappa_\lambda$ can be as large as 25. 
The $b\rightarrow s \gamma$ bound, in green, is very effective in reducing the splitting between $m_{H^\pm}$ and $m_{A}$ to at most 250 GeV. It also restricts the splitting between $m_{H}$ and $m_{A}$, but to a lesser extent. However, in relation to $\kappa_{\lambda}$, this bound alone allowed for values as large as 25.
Finally, it should be noticed that the allowed region after imposing collider bounds, in orange, is larger than the allowed region when considering $b \rightarrow s \gamma$. However, collider bounds are the most restrictive to the maximal value of $\kappa_\lambda$, by enforcing $s_{\beta-\alpha}$ to be close to one. As a result, the region with large $\tan\beta$ is also excluded, which was responsible for the highest values for $\kappa_{\lambda}$. 

\begin{figure}[h!]
    \centering
    \includegraphics[scale=0.8]{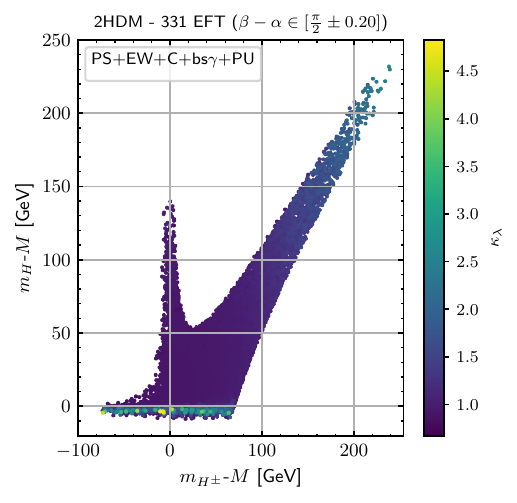}
    \includegraphics[scale=0.8]{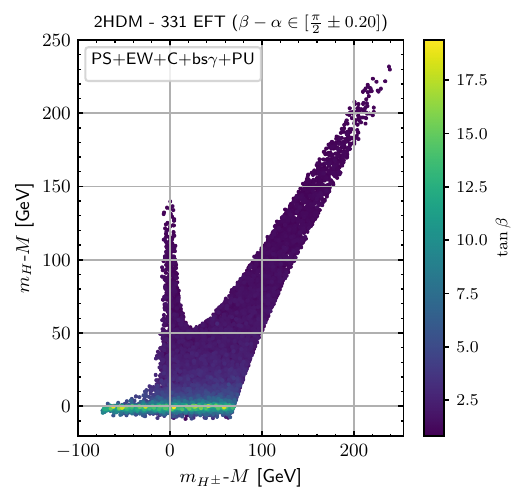}
    \caption{Parameter space of mass difference for points that comply with all bounds, as a function of the value of $\kappa_\lambda$ (left) or $\tan{\beta}$ (right).}
    \label{fig:kappa_final}
\end{figure}

\begin{figure}[h!]
    \centering
    \includegraphics[scale=0.8]{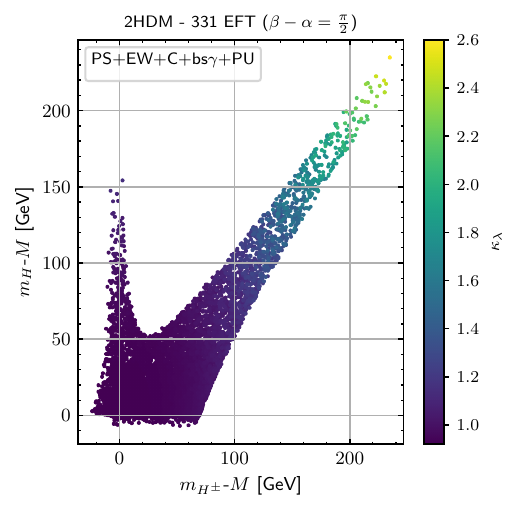}
    \caption{Parameter space of mass difference for points that comply with all bounds, as a function of the value of $\kappa_\lambda$, imposing the alignment condition ($s_{\beta-\alpha}=1$).}
    \label{fig:kappa_final-A}
\end{figure}

The final allowed region, which complies with all bounds simultaneously, is shown in~\cref{fig:kappa_final} (left). 
As anticipated, we observe larger values of $\kappa_\lambda$ in the region where both $m_H$ and $m_{H^\pm}$ have a large splitting with $M$. At first glance, it is surprising to also see large values of $\kappa_\lambda$ in the lower left part of the parameter space, that is, for $m_H\sim m_{H^\pm}\sim M$. As we show in~\cref{fig:kappa_final} (right), this region is exactly where $\tan{\beta}$ is the largest.
To show the impact of the out-of-alignment condition on the maximal value of $\kappa_\lambda$, we show the same results but enforcing $s_{\beta-\alpha}=1$ in~\cref{fig:kappa_final-A}.
Note that the maximal value of $\kappa_\lambda$ is reduced~\footnote{When including two-loop corrections, which are known only at the alignment regime, it is still possible to attain values for $\kappa_\lambda\sim 3.5$. See~\cref{ap:two-loop}.} and now the region where $m_H\sim M$ no longer provides sizable values for the trilinear Higgs coupling. This shows that even with the very restrictive experimental constraints to $s_{\beta-\alpha}$, and thus, allowing only very small deviations to unity, it is possible to significantly increase the trilinear coupling in this situation. For completeness, we discuss in~\cref{ap:2HDM} the allowed region for a 2HDM Type I and II, as opposed to the 3-3-1 EFT considered in this work.

\begin{figure}[h!]
    \centering
    \includegraphics[scale=0.75]{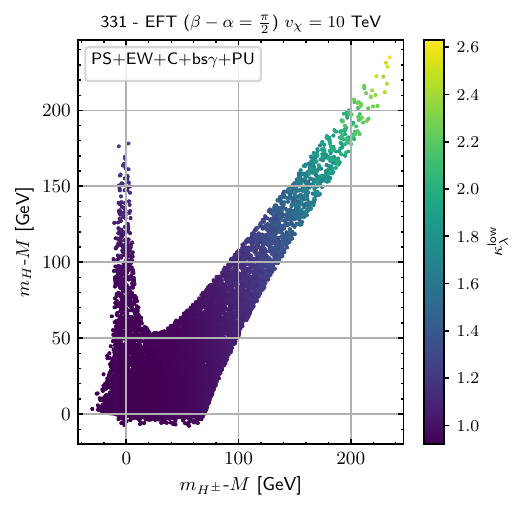}
    \includegraphics[scale=0.75]{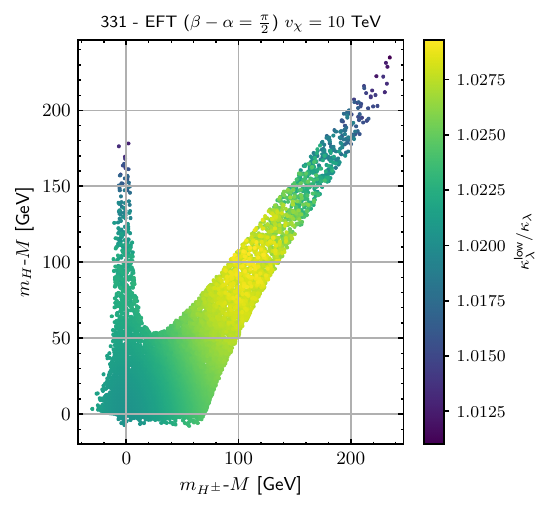}
    \caption{Parameter space of mass difference for points that comply with all bounds as function of the value of $\kappa_\lambda$. On the left, we consider all the EFT matching terms up to dimension 6, which we define by $\kappa_\lambda^{\rm{low}}$. On the right, we compare $\kappa_\lambda^{\rm{low}}$, computed with $v_\chi=10\;\rm{TeV}$, and $\kappa_\lambda$ (for which $v_\chi\sim100\;\rm{TeV}$).}
    \label{fig:kappa_final-331-dim6}
\end{figure}

\begin{figure}[h!]
    \centering
    \includegraphics[scale=0.75]{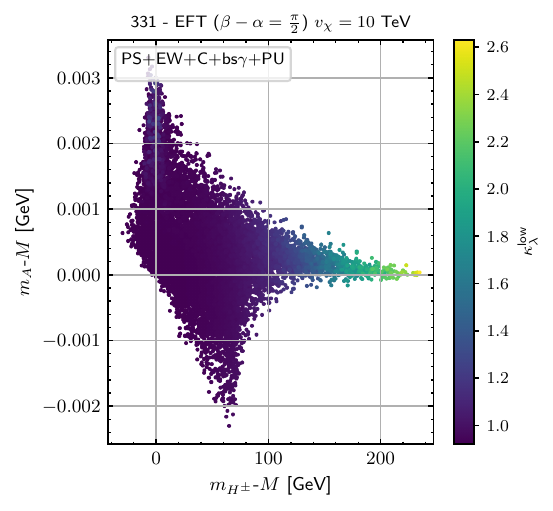}
    \includegraphics[scale=0.75]{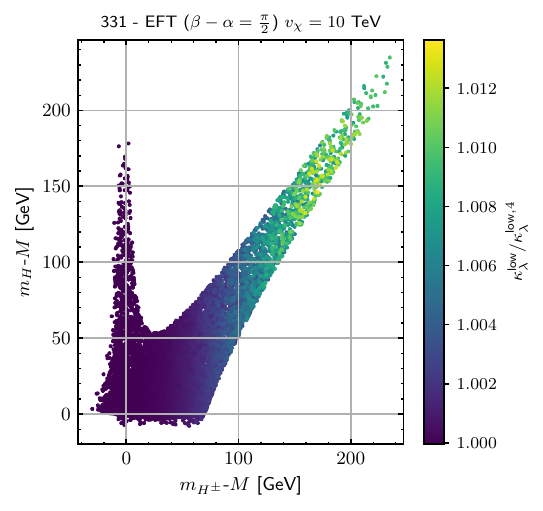}
    \caption{Parameter space of mass difference for points that comply with all bounds as function of the value of $\kappa_\lambda$. On the left, we consider all the EFT matching terms up to dimension 6, defined by $\kappa_\lambda^{\rm{low}}$. On the right, we compare the same full matching of the 331 - EFT with the case that dimension 6 operators are neglected.}
    \label{fig:kappa_final-331-dim6_comp}
\end{figure}

The previous analysis was performed for a very large value of $v_{\chi}$, in order to extract the low-energy consequences, in particular the maximal attainable $\lambda_{hhh}$, for this pessimistic scenario (where none of the heavier spectrum will be detected, and their imprint on low-energy observables is negligible). If the heavy scale is lower, we expect some influence on the value of $\lambda_{hhh}$, as we mentioned before. After applying the same constraints as before, but considering the full dependence on the heavy scale and the inclusion of dimension-six terms, we obtain the allowed parameter space depicted in \cref{fig:kappa_final-331-dim6}. The figure on the left should be compared against \cref{fig:kappa_final-A}. As can be seen, the maximum value for $\kappa_{\lambda}$ is similar; however, it is possible to attain higher values for lower mass splittings. This can be seen by inspecting the figure on the right. Regarding the mass split between $M$ and $m_{A}$ for $v_{\chi}=10\;\rm{TeV}$, it can be seen in \cref{fig:kappa_final-331-dim6_comp} (left). It is instructive to notice that $\kappa_{\lambda}$ is maximum for $M\sim m_{A}$. In the right, we depict the ratio between the case with full matching plus dimension six against the case where dimension-six terms are neglected. We notice that the inclusion of the latter amounts to $1 \%$ or lower difference. We emphasize that a complete treatment of dimension-six operators, including RGE running and operator-mixing, is beyond the scope of our work. Our main aim is to provide an estimate of their influence, when the heavy scale is still at the LHC energy domain. Nevertheless, by assuming the same coefficients and just plugging in the large log ($\log(v_{\chi}^{2}/v_{\rm{ew}}^{2})\sim 3$), the inclusion of dimension-six terms would still amount to a correction at the percent level.

\begin{figure}[h!]
    \centering
    \includegraphics[scale=0.8]{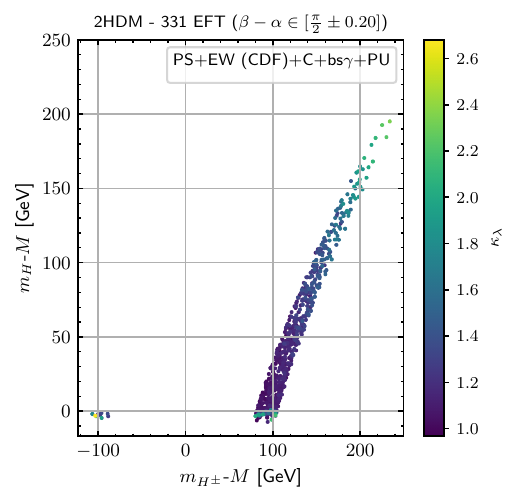}
    \caption{Parameter space of mass difference for points that comply with all bounds as function of the value of $\kappa_\lambda$. We, however, use the STU observables calculated using the CDF measurement of $m_W$, as opposed to the PDG combined fit.}
    \label{fig:kappa_final-CDF}
\end{figure}

Finally, if we instead consider the measurement from CDF for the W-mass, we assume that this measurement does not alter significantly other well-established parameters, namely the Fermi constant, the Yukawa couplings, and the SM vacuum expectation value. Therefore, the only significant impact of this is in the STU parameters, as already discussed. We then obtain the result shown in~\cref{fig:kappa_final-CDF}.
We see that, even though the allowed region for the modified STU parameters considering the CDF measurement, shown in~\cref{fig:stu_kappa_comp}, allows a slightly higher value of $\kappa_\lambda$ compared to the PDG measurement, shown in~\cref{fig:stu_kappa}, these bigger values of $\kappa_\lambda$ vanish when other bounds are imposed, mostly collider bounds.

In conclusion, we find that combining all the most important constraints, there are two different conditions to maximize $\kappa_\lambda$ in this model, a small mass splitting with a relatively large $\tan{\beta}$ (order of $\tan{\beta}\sim 10$), or with large mass splittings with $M$. The former condition is mostly restricted by collider searches and to a lesser degree by EW observables, yielding the largest value for $\kappa_\lambda$, while the latter is mostly restricted by PU, bs$\gamma$ and collider bounds. 

Furthermore, with a maximal value of order $4.5$, we conclude that this parameter space can be further constrained and explored with the next phase of the LHC, with a higher luminosity~\cite{Gianotti:2002xx,Chang:2019ncg,Cepeda:2019klc,ATL-PHYS-PUB-2022-053} and also with the potential construction of next-generation colliders, such as the Future Circular Collider (FCC)~\cite{Arkani-Hamed:2015vfh,Baglio:2015wcg,Contino:2016spe}, for example.

\section{Conclusions}
\label{sec:conclusions}
Given the non-observation of mass resonances at the LHC, the path to BSM models is still unclear. In particular, it is possible that the scale of New Physics is higher than the one probed at the LHC. In this scenario, the effective field theory framework can be particularly useful, by parameterizing the effects of UV models in low-energy observables. In this work, we focus on the 3-3-1 class of models as UV completion, presenting a general parameter scan of the low-energy limit of such models. With the heavy degrees of freedom integrated out, we obtain an effective field theory (3-3-1 EFT) at one-loop matching, which can be mapped into a kind of 2HDM EFT Type II. We show that some of the terms of the general, renormalizable 2HDM potential, namely ($\Lambda_5$, $\Lambda_6$ and $\Lambda_7$) are naturally suppressed by $v_{\rm{SM}}/v_{\chi}$, where $v_\chi$ represents the heavy scale. The effective model also features a different structure for the Yukawa couplings between generations; however, by restricting to the third family, the couplings can be exactly mapped to the 2HDM EFT Type II. 

We performed a scan in the parameter space of the theory, imposing bounds of perturbativity, vacuum stability, perturbative unitarity, $S$, $T$, and $U$ precision parameters at one-loop, collider known resonances, and direct scalar searches, as well as $B\rightarrow{} X_s \gamma$ decay rate, evaluating the potential impact of each of these bounds on the maximal value of the trilinear Higgs coupling. We find that the collider constraints are the most effective in limiting the value of $\kappa_\lambda$, mostly due to its impact on the bounds in $s_{\beta-\alpha}$, $\tan{\beta}$, and $m_{\Phi}-M$ ($\Phi = H$, $A$, $H^{\pm}$). In this work, we show that even for small deviations of the alignment condition, $s_{\beta-\alpha}\approx 0.995$, allowed by experiments, the impact of $s_{\beta-\alpha}$ and $\tan{\beta}$ can be significant, even generating the highest values of $\kappa_\lambda$. We also present an approximate formula for $\kappa_\lambda$ at one-loop, valid for the non-alignment scenario, which we confront with the full calculation, obtaining a good agreement. Finally, we get a maximal value of $\kappa_\lambda \approx 4.5$ for the 3-3-1 EFT, which means that it is possible to explore this feature of the model in the next few years after the upcoming upgrades in the LHC and future colliders.

As possible extensions to our work, we point out a comprehensive analysis of the trilinear higgs coupling in the context of the 2HDM EFT by itself, not connected to any UV completion in particular. In this context, a full analysis of the influence of dimension-six terms seems to be a promising avenue, which could be achieved in similarity to previous analyses performed in the context of the SMEFT. In addition, the computation and subsequent inclusion of the RGE effects in the 2HDM EFT is desirable, in particular if the UV cut-off proves to be very far from the electroweak scale.

\acknowledgments

A.L.C acknowledges enlightening discussions with Mikael Chala during the earlier stages of this work. A.L.C is supported by a postdoctoral fellowship from the Postdoctoral Researcher Program - Resolution GR/Unicamp No. 33/2023. LJFL is thankful for the support of CAPES under grant No. 88887.613742/2021-00.

\appendix

\section{Type I and Type II models}
\label{ap:2HDM}

In this section, our aim is to compare the different 2HDM types, more specifically Types I and II and the EFT arising from the 3-3-1 model. The same conclusions can be easily extended to the lepton-specific and flipped ones, since the Yukawa coupling to leptons does not play a significant role in this analysis. Therefore, we can immediately map the Lepton-Specific to Type I and the Flipped to Type II.

\begin{figure}[h!]
    \centering
    \includegraphics[scale=0.8]{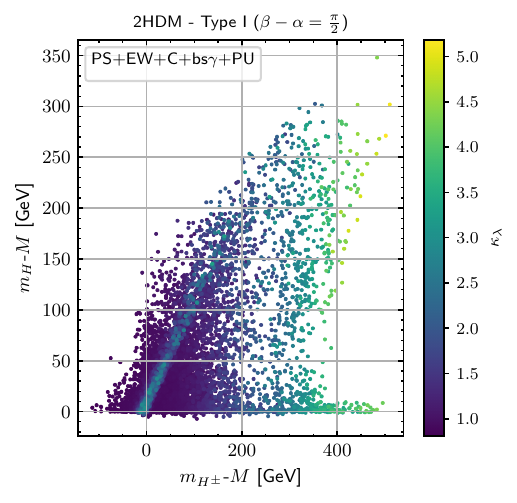}
    \includegraphics[scale=0.8]{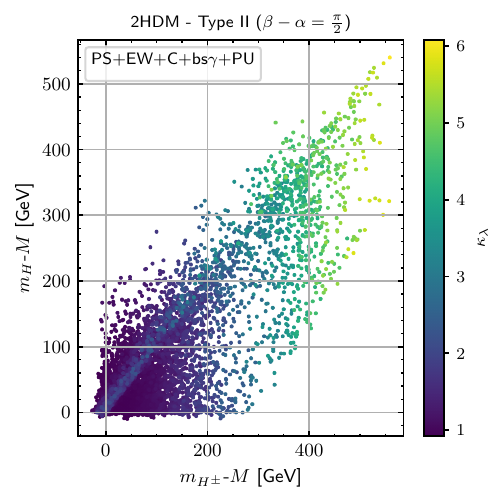}
    \caption{Parameter space of mass difference for points that comply with all bounds as function of the value of $\kappa_\lambda$ for 2HDM types I (left) and II (right), under the alignment condition.}
    \label{fig:kappa_final-I-and-II}
\end{figure}

\begin{figure}[h!]
    \centering
    \includegraphics[scale=0.8]{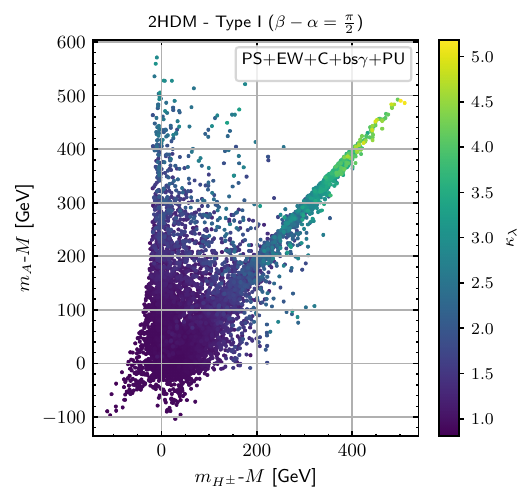}
    \includegraphics[scale=0.8]{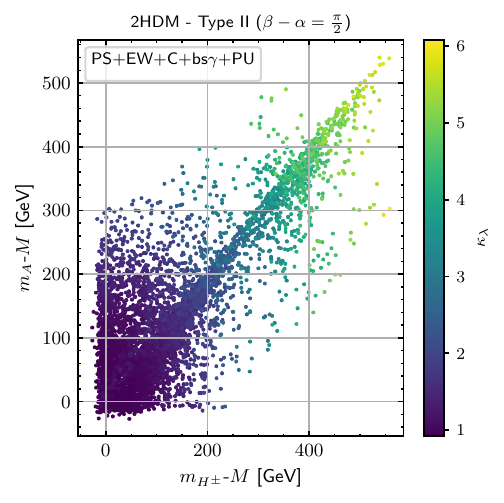}
    \caption{Parameter space of mass difference for points that comply with all bounds as function of the value of $\kappa_\lambda$ for 2HDM types I (left) and II (right), under the alignment condition.}
    \label{fig:kappa_final-I-and-II-1}
\end{figure}

The results of our scan, for the alignment scenario, are shown in~\cref{fig:kappa_final-I-and-II,fig:kappa_final-I-and-II-1}.

We find that, in the alignment limit, there is no significant difference between Type I and Type II. The trilinear coupling can be slightly bigger for Type II, however, it is not possible to assert with certainty that this is not due to the finite amount of points considered in the scan. 

Since these models, in general, have $\lambda_5\neq 0$, they have one extra free parameter which can increase the value of $\kappa_\lambda$. In fact, compared to~\cref{fig:kappa_final-A}, it is noticeable that they allow larger values of $\kappa_\lambda$ compared to the 3-3-1 EFT. This extra free parameter, in the form of $m_A-M$, is shown in~\cref{fig:kappa_final-I-and-II-1}.

It is possible to see that a large splitting between $m_A$ and $M$ significantly increases the value of $\kappa_\lambda$, especially when such splitting also occurs for $m_{H^\pm}$. This gives this particular V-shaped format in the parameter space of~\cref{fig:kappa_final-I-and-II-1}, where the ``extremities of the V'' reach the largest values of the trilinear coupling.
On the other hand, from~\cref{fig:kappa_final-I-and-II}, it is possible to see that a large splitting of $m_H$ from $M$ while keeping $m_A$, $m_{H^\pm}$ in the same scale of $M$ is not attainable, nor there is a large impact of $m_H-M$ on the value of $\kappa_\lambda$, compared to the other splittings. Still, the conclusion is the same as before, the largest possible values for the trilinear coupling are obtained when $m_\Phi \gg M$ and the physical masses are more or less in the same order of magnitude.

\begin{figure}[h!]
    \centering
    \includegraphics[scale=0.8]{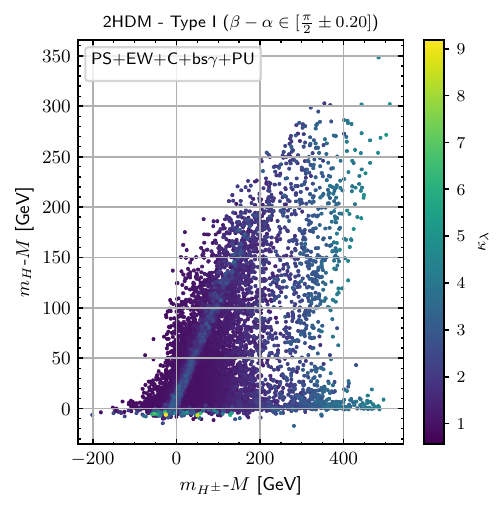}
    \includegraphics[scale=0.8]{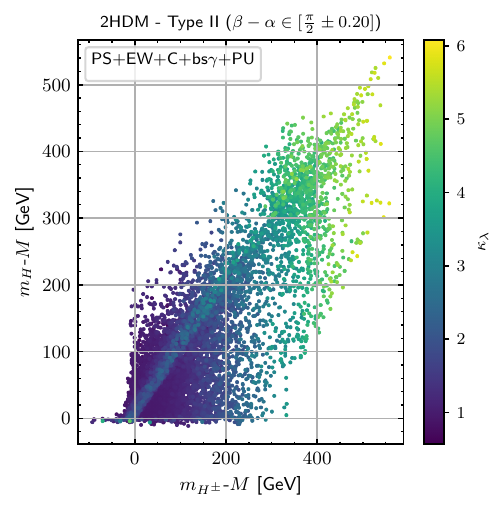}
    \caption{Parameter space of mass difference for points that comply with all bounds as function of the value of $\kappa_\lambda$ for 2HDM types I (left) and II (right).}
    \label{fig:kappa_final-I-and-II-NA}
\end{figure}

\begin{figure}[h!]
    \centering
    \includegraphics[scale=0.8]{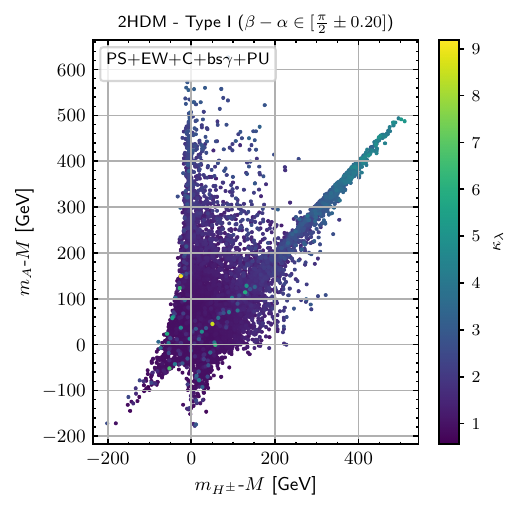}
    \includegraphics[scale=0.8]{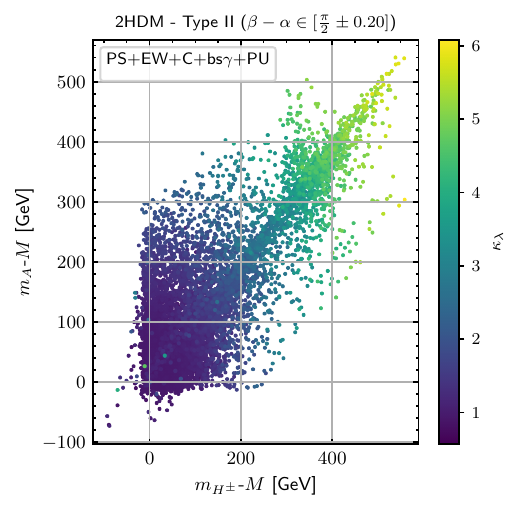}
    \caption{Parameter space of mass difference for points that comply with all bounds as function of the value of $\kappa_\lambda$ for 2HDM types I (left) and II (right).}
    \label{fig:kappa_final-I-and-II-1-NA}
\end{figure}

In~\cref{fig:kappa_final-I-and-II-NA,fig:kappa_final-I-and-II-1-NA} we show how this picture changes if the restriction on the alignment condition is lifted. 

Compared to~\cref{fig:kappa_final-I-and-II,fig:kappa_final-I-and-II-1}, it is immediately evident that the maximal value of the trilinear coupling is approximately twice the maximum value when enforcing the alignment condition for Type I 2HDM. This is not unexpected, since it is well known in the literature~\cite{ATLAS:2024lyh} that collider constraints are much more severe for Type II compared to Type I. Indeed, since slightly smaller $s_{\beta-\alpha}$ and larger values of $\tan{\beta}$ are allowed, the impact of these parameters can be seen more strongly in $\kappa_\lambda$.

Now, for Type II, compared to the 3-3-1 EFT in the alignment scenario (\cref{fig:kappa_final-A}), one sees that the addition of $M$ as a free parameter (\cref{fig:kappa_final-I-and-II,fig:kappa_final-I-and-II-1}) allows $\kappa_{\lambda}$ to be twice as large. When one further considers out-of-alignment Type II (\cref{fig:kappa_final-I-and-II-NA,fig:kappa_final-I-and-II-1-NA}), the maximal value for the trilinear Higgs coupling is roughly the same. For the 3-3-1 EFT, on the other hand, lifting the restriction on the alignment condition allows significantly larger values for $\kappa_{\lambda}$ (\cref{fig:kappa_final}), of the order of those found in Type II.

\section{Two-loop calculation}
\label{ap:two-loop}

In this section, we discuss the influence of two-loop corrections for our findings. As already stated, those are known only for the alignment regime~\cite{Braathen:2019pxr,Braathen:2019zoh}. In these references, $\kappa_{\lambda}$ is expressed as
\begin{align}
    \kappa_{\lambda}=\kappa_{\lambda}^{(0)}+\left(\frac{1}{16\pi^2}\right)\kappa_{\lambda}^{(1)}+\left(\frac{1}{16\pi^2}\right)^{2}\kappa_{\lambda}^{(2)}+...
\end{align}

The explicit formula for $\kappa_{\lambda}^{(2)}$ is given for the $\overline{\mbox{MS}}$ regularization scheme, together with all the ingredients to obtain $\kappa_{\lambda}^{(2)}$ at the on-shell (OS) scheme as well. Following the required steps, we obtained $\kappa_{\lambda}^{(2)}$ in the OS, and checked that the dependence on the regularization scale $\mu^2$ cancels as expected.

Once $\kappa_{\lambda}^{(2)}$ is known on the OS scheme, we can study how our findings for the 3-3-1 EFT in the alignment scenario are modified. In order to make this comparison, we will define $\kappa_{\lambda}^{\rm{NLO}}$ to be the trilinear higgs couplings modifier evaluated up to one-loop order, which we have considered in this manuscript up to this point. When two-loop corrections are also included, we obtain $\kappa_{\lambda}^{\rm{NNLO}}$. 

\begin{figure}[h!]
    \centering
    \includegraphics[scale=0.8]{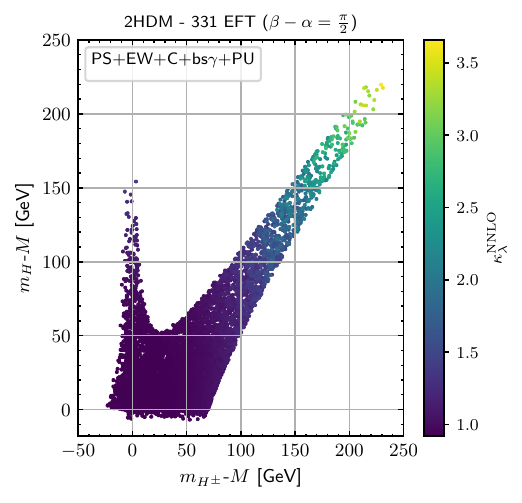}
    \includegraphics[scale=0.8]{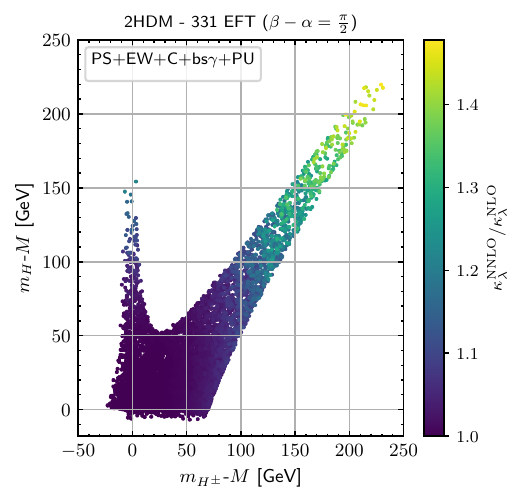}
    \caption{Parameter space of mass difference for points that comply with all bounds as function of the value of $\kappa_\lambda$ at NNLO (left) and the ratio of NNLO by NLO (right) for 3-3-1 EFT.}
    \label{fig:kappa_final-331-twoloop}
\end{figure}

In \cref{fig:kappa_final-331-twoloop} (left) we show the effect of including two-loop corrections for $\kappa_{\lambda}$. As can be easily seen, larger values are achieved for $m_{A}\ll m_{H}\sim m_{H^{\pm}}$, which can reach 3.5. This should be compared against \cref{fig:kappa_final-A}, where the same pattern emerges but $\kappa_{\lambda}$ is at most 2.4. Thus, when including the two-loop corrections, one can obtain values up to $45\%$ larger. We quantify this feature by showing in \cref{fig:kappa_final-331-twoloop} (right) the ratio between the NNLO and NLO contributions. Not only does the region $m_{A}\ll m_{H}\sim m_{H^{\pm}}$ allow the highest values for $\kappa_{\lambda}^{\rm{NLO}}$, but it also maximizes the contribution of two-loop corrections. 

We have also performed an analysis for the 2HDM Types I and II, finding similar results from~\cite{Bahl:2022jnx}.

\section{Dimension six operators and one-loop matching}
\label{ap:one-loop}

In this appendix, we provide explicit results regarding the matching at one-loop order. We also consider dimension-six operators that encompass contributions from the EFT to the trilinear higgs coupling, and perform the one-level matching for those as well. 

We start with the one-loop matching for the coefficients $\Lambda_5$, $\Lambda_6$, $\Lambda_7$ that are naturally suppressed at tree-level matching. Regarding the contribution of heavy scalars, we first resort to \cref{eq:Vscalar2} where the first gauge breaking was already performed and all heavy scalars are mass eingenstates. Under these conditions, it is possible to employ the code \textit{Matchete}~\cite{FuentesMartin:2022jrf} to obtain the one-loop matching to $\Lambda_{i}$ from all heavy scalars. With regard to the heavy fermions, we need to use \cref{eq:fermionQ,eq:fermionE}, where the first breaking was already performed. However, the heavy fermions ($D,S,T,E_i$) are not mass eingenstates yet. This can be easily accomplished by going to the basis where $y_{2}^{D}=y_{1}^{S}=0$ and $y_{mn}^{E}$ is a diagonal matrix. Once again, after these conditions are met, we can resort to \textit{Matchete} to perform the one-loop matching from the heavy fermions. 

Finally, for the heavy gauge bosons, it is more involved to employ the \textit{Matchete} code. It requires that the gauge group for the underlying model is defined in the first place, handling the covariant derivatives, and consequently the couplings between gauge bosons and the other particles of the model, consistently. At this stage, no gauge breaking was performed yet, implying that all gauge bosons are massless and that no EFT
from them can be devised. Given this obstacle, we will refrain from computing the explicit one-loop matching from the heavy gauge bosons. Nevertheless, it is possible to convince ourselves that their contribution should be sub-dominant. As already emphasized, they can only affect one-loop matching. Ultimately, since we are interested in diagrams with four-external legs containing light scalar fields only, they will contribute to one-loop diagrams such as boxes. In this case, the couplings between the light scalars and the heavy gauge boson is proportional to the extra gauge coupling, which is necessarily below unity (we are considering a weakly interacting theory). Therefore, their contribution should be of the form
\begin{equation}
    \mathcal{M}_{g}\sim g^{4}\int_{k}\frac{1}{(k^{2}-M^{2})^{4}}\sim \frac{g^{4}}{M^{4}}
\end{equation}
which is more suppressed than the main contribution from fermions, that goes as $1/M^{2}$. Even if we consider triangles instead of boxes, the contribution from heavy gauge bosons will be proportional to powers of the gauge coupling. This is below unity, while the Yukawa coupling can be of order one, rendering the fermionic contribution more important. A similar argument can be applied to the heavy scalars contribution, where the triple scalar couplings can be large.

Since the results are lengthy, we adopt some simplifications. We will consider that the heavy particles (scalars, fermions) are degenerate in mass ($M_{X}$) for simplicity, as well as set the renormalization scale at $M_{X}$. The general result (also for couplings $\Lambda_{1},\Lambda_{2},\Lambda_{3},\Lambda_{4}$) is provided in the auxiliary notebook\footnote{\url{https://github.com/LeoFerreira8/hhh---General}}, while the simplified result for $\Lambda_{5},\Lambda_{6},\Lambda_{7}$ is given by 

\begin{align}
    \frac{\Lambda_{5}}{2} = &-\frac{f^2}{M_X^2}\Bigg[1 +   \frac{1}{8 \pi^2} ( \zeta_{12} + \zeta_{13} + \zeta_{23} + \lambda_1 + \lambda_{12} - \lambda_{13} + \lambda_2 - \lambda_{23} + 6 \lambda_3 \nonumber\\
    & \qquad\qquad\qquad\qquad  -6 \,\bar{y}_{D}{y}_{D}-6\, \bar{y}_{S}{y}_{S}-6\, \bar{y}_{T}{y}_{T}-2\, {\rm{Tr}}[\bar{y}_{E}{y}_{E}])\Big]\\
        \Lambda_6 =  &\frac{2 \sqrt{2} f \, v_\chi \, \lambda_{13}}{M_X^2}\Bigg\{1 + \frac{1}{8\pi^2} \Bigg[ 4 \zeta_{13} + 4 \zeta_{23}  + 6 \lambda_1  + 2 \lambda_{12}  - 4 \lambda_{13}  + \lambda_{12} \frac{\lambda_{23}}{\lambda_{13}} + \zeta_{12} \left(2  + \frac{\lambda_{23}}{\lambda_{13}}\right)  \nonumber\\
       & \qquad\qquad\qquad\qquad\qquad + 24  \lambda_3  -12\,\bar{y}_{D}{y}_{D}+12\, \bar{y}_{S}{y}_{S}+12\, \bar{y}_{T}{y}_{T}+ 4\,{\rm{Tr}}[\bar{y}_{E}{y}_{E}]\nonumber\\
       & \qquad\qquad\qquad\qquad\qquad - \frac{ 4 M_X
       }{v_{\chi}} \Big[3\, {\rm{Re}} (y_{D}+y_{S}+y_{T}) +  {\rm{Tr}}(y_{E}) \Big]\Bigg]\Bigg\}\\
        \Lambda_7 =  &\frac{2 \sqrt{2} f \, v_\chi \, \lambda_{23}}{M_X^2}\Bigg\{1 + \frac{1}{8\pi^2} \Bigg[ 4 \zeta_{13} + 4 \zeta_{23}  + 6 \lambda_2  + 2 \lambda_{12}  - 4 \lambda_{23}  + \lambda_{12} \frac{\lambda_{13}}{\lambda_{23}} + \zeta_{12} \left(2  + \frac{\lambda_{13}}{\lambda_{23}}\right)  \nonumber\\
       & \qquad\qquad\qquad\qquad\qquad + 24  \lambda_3  -12\,\bar{y}_{D}{y}_{D}+3\, \bar{y}_{S}{y}_{S}+\, \bar{y}_{T}{y}_{T}+ {\rm{Tr}}[\bar{y}_{E}{y}_{E}]\nonumber\\
       & \qquad\qquad\qquad\qquad\qquad - \frac{ 4 M_X
       }{v_{\chi}} \Big[3\, {\rm{Re}} (y_{D}+y_{S}+y_{T}) +  {\rm{Tr}}(y_{E}) \Big]\Bigg]\Bigg\}
       \end{align}

Once the corrections to $\Lambda_{i}$ are known, we can estimate their influence on the trilinear Higgs coupling. To this end, we obtain the tree-level value for $\kappa_{\lambda}$ in the 2HDM with non-null $\Lambda_{5,6,7}$ and replace them by the one-loop matching formulas. Once again, the general result is provided in the auxiliary notebook.

Since we are considering the 2HDM EFT, dimension-six operators (and higher) are also present, which may contribute to the trilinear higgs coupling. However, a full analysis is beyond the scope of this work. Our main aim is to estimate the impact these operators may have, after the matching to the 3-3-1 model. In order to do that, we considered all operators containing six scalar fields in the 2HDM EFT, which we collect in \cref{tab:dimension6}. We performed one-loop matching in the symmetry basis, then realized the gauge breaking, and finally resorted to the mass eingenstate basis. In the latter, we extracted the correction to the trilinear higgs coupling. The general result in terms of the scalar potential parameters can be found in the auxiliary notebook. We also provide the result in terms of the mass of the light scalars and the mixing angles for the simplified case where $\lambda_{i3}\sim0$. It is still lengthy, thus we refrain from presenting it here. In the alignment limit, we obtain the result given in \cref{eq:hhh_one}. 

\begin{table}[t]
\centering
 \begin{tabular}{|l l l||} 
    \hline
   $\mathcal{O}_{\Phi}^{(11)(11)(11)} = (\Phi_1^{\dagger} \Phi_1)^3 $ & $\mathcal{O}_{\Phi}^{(11)(12)(21)} = (\Phi_1^{\dagger} \Phi_1) (\Phi_1^{\dagger} \Phi_2) (\Phi_2^{\dagger} \Phi_1)$ & 
    $\mathcal{O}_{\Phi}^{(12)(21)(21)} = (\Phi_1^{\dagger} \Phi_2)^2 (\Phi_2^{\dagger} \Phi_1)^2$ \\ $\mathcal{O}_{\Phi}^{(11)(11)(22)} = (\Phi_1^{\dagger} \Phi_1)^2(\Phi_2^{\dagger} \Phi_2)$ &
     $\mathcal{O}_{\Phi}^{(11)(22)(22)} = (\Phi_1^{\dagger} \Phi_1)(\Phi_2^{\dagger} \Phi_2)^2$ & $\mathcal{O}_{\Phi}^{(22)(22)(22)} = (\Phi_2^{\dagger} \Phi_2)^3$ \\
     \hline
    \end{tabular}
    \begin{tabular}{|l l||} 
    \hline
     $\mathcal{O}_{\Phi}^{(11)(21)(21)} = (\Phi_1^{\dagger} \Phi_1) (\Phi_2^{\dagger} \Phi_1)^2+ h.c.$ &
     $\mathcal{O}_{\Phi}^{(11)(11)(21)} = (\Phi_1^{\dagger} \Phi_1)^2 (\Phi_2^{\dagger} \Phi_1)+ h.c.$ \\
     $\mathcal{O}_{\Phi}^{(12)(21)(21)} = (\Phi_1^{\dagger} \Phi_2) (\Phi_2^{\dagger} \Phi_1)^2+ h.c.$ &
     $\mathcal{O}_{\Phi}^{(21)(21)(22)} = (\Phi_2^{\dagger} \Phi_1)^2 (\Phi_2^{\dagger} \Phi_2)+ h.c.$ \\
     $\mathcal{O}_{\Phi}^{(11)(21)(22)} = (\Phi_1^{\dagger} \Phi_1) (\Phi_2^{\dagger} \Phi_1)(\Phi_2^{\dagger} \Phi_2)+ h.c.$ &
     $\mathcal{O}_{\Phi}^{(21)(22)(22)} = (\Phi_2^{\dagger} \Phi_1) (\Phi_2^{\dagger} \Phi_2)^2+ h.c.$  \\
    \hline
    \end{tabular}
    \caption{Dimension-six operators that contribute to the trilinear higgs coupling.}
\label{tab:dimension6}
\end{table}

\begin{align}
    \label{eq:hhh_one}   
\lambda_{hhh}^{(6)}=&\frac{120\, f^2 m_{h}^2 t_{\beta}^2 v}{M_X^4 \left(1+t_{\beta}^2\right)^2}\nonumber\\&-\frac{5m_h^6}{16  \pi^2 M_X^2 v^3}\left[  1- \frac{3 (m_A^2 - m_{H_{\pm}}^2)}{m_h^2} 
+ \frac{6 (m_A^2 - m_{H_{\pm}}^2)^2}{m_h^4} 
- 4 \frac{(m_A^2 - m_{H_{\pm}}^2)^3}{m_h^6}\right]
 \,.
\end{align}

\bibliographystyle{JHEP}
\bibliography{draft}

\end{document}